\documentclass[letter]{IEEEtran}
\IEEEoverridecommandlockouts

\usepackage{kotex}
\usepackage{url}
\usepackage{cite}
\usepackage{lipsum} 
\usepackage{graphicx}
\usepackage{amsmath, amsfonts, amssymb}
\usepackage{xcolor}
\usepackage{booktabs, multirow, array}
\ifCLASSOPTIONcompsoc
    \usepackage[caption=false, font=normalsize, labelfont=sf, textfont=sf]{subfig}
\else
\usepackage[caption=false, font=footnotesize]{subfig}

\usepackage[ruled, linesnumbered]{algorithm2e}
\usepackage{etoolbox}
\makeatletter
\patchcmd{\@algocf@start}
  {-1.5em}
  {0pt}
  {}{}
\makeatother

\newtheorem{theorem}{Theorem}

\newtheorem{lemma}{Lemma}

\usepackage{stackengine}
\DeclareMathOperator*{\defeq}{\mathrel{\ensurestackMath{\stackon[1pt]{=}{\scriptscriptstyle\Delta}}}}

\newcommand{\Rom}[1]{\MakeUppercase{\romannumeral #1}}

\begin{document}
\def\cA{{\mathcal{A}}} \def\cB{{\mathcal{B}}} \def\cC{{\mathcal{C}}} \def\cD{{\mathcal{D}}}
\def\cE{{\mathcal{E}}} \def\cF{{\mathcal{F}}} \def\cG{{\mathcal{G}}} \def\cH{{\mathcal{H}}}
\def\cI{{\mathcal{I}}} \def\cJ{{\mathcal{J}}} \def\cK{{\mathcal{K}}} \def\cL{{\mathcal{L}}}
\def\cM{{\mathcal{M}}} \def\cN{{\mathcal{N}}} \def\cO{{\mathcal{O}}} \def\cP{{\mathcal{P}}}
\def\cQ{{\mathcal{Q}}} \def\cR{{\mathcal{R}}} \def\cS{{\mathcal{S}}} \def\cT{{\mathcal{T}}}
\def\cU{{\mathcal{U}}} \def\cV{{\mathcal{V}}} \def\cW{{\mathcal{W}}} \def\cX{{\mathcal{X}}}
\def\cY{{\mathcal{Y}}} \def\cZ{{\mathcal{Z}}} \def\cz{{\mathcal{z}}}

\title{Sparsely Pre-transformed Polar Codes for Low-Latency SCL Decoding}


\author{   
Geon Choi, \IEEEmembership{Student Member, IEEE} and
 Namyoon Lee, \IEEEmembership{Senior Member, IEEE}
\thanks{
Geon Choi is with the Department of Electrical Engineering, POSTECH, Pohang 37673, South Korea
(e-mail: \mbox{geon.choi@postech.ac.kr}).
}
\thanks{
Namyoon Lee is with the School of Electrical Engineering, Korea University, Seoul 02841, South Korea (e-mail: \mbox{namyoon@korea.ac.kr}).
}
\thanks{
This work was presented in part at IEEE ISIT 2024 \cite{spp-isit}.}

}

\maketitle

\begin{abstract} 
Deep polar codes, employing multi-layered polar kernel pre-transforms in series, are recently introduced variants of pre-transformed polar codes. These codes have demonstrated the ability to reduce the number of minimum weight codewords, thereby closely achieving finite-block length capacity with successive cancellation list (SCL) decoders in certain scenarios. However, when the list size of the SCL decoder is small, which is crucial for low-latency communication applications, the reduction in the number of minimum weight codewords does not necessarily improve decoding performance. To address this limitation, we propose an alternative pre-transform technique to enhance the suitability of polar codes for SCL decoders with practical list sizes. Leveraging the fact that the SCL decoding error event set can be decomposed into two exclusive error event sets, our approach applies two different types of pre-transformations, each targeting the reduction of one of the two error event sets. Extensive simulation results under various block lengths and code rates have demonstrated that our codes consistently outperform all existing state-of-the-art pre-transformed polar codes, including CRC-aided polar codes and polarization-adjusted convolutional codes, when decoded using SCL decoders with small list sizes.

\end{abstract}


\section{Introduction}

The demand for ultra-reliable low-latency communication (URLLC) persists in next-generation wireless communication systems. URLLC aims to deliver extremely high-speed data packets with a very low packet error rate within a very short timeframe \cite{Wang-6G-Survey,   david-6G-URLLC, 6G-next-frontier}. To achieve these goals, the development of cutting-edge channel coding technology is essential. This technology must be capable of exhibiting excellent error-correcting performance and fast decodability within a finite blocklength regime \cite{shirvanimoghaddam19, yue23, stephan-welcome-chance, BOSS2,BOSS-URLLC,BOSS3}.

Pre-transformed polar codes are a variant of polar codes that involve applying an upper-triangular pre-transform before the polar coding process. This pre-transformation has been shown to enhance the weight spectrum of the codes by either reducing the number of minimum weight codewords or increasing the minimum distance. Numerous pre-transformation techniques have been proposed in the literature \cite{Niu-CA-polar, Trifonov-polar-dynamic-frozen, Wang-PCC-polar, Zhang-PC-polar-Huawei, arikan-pac, Gelincik-polar-row-merging, zunker-row-merged, deep-polar-construction-A, deep-polar-gc-wkshps}. Most of these methods aim to improve the code weight spectrum and are particularly effective when using maximum likelihood (ML) decoders or successive cancellation list (SCL) decoders with large list sizes \cite{Chiu-PAC-construction, Miloslavskaya-design-short-polar, zunker-enumeration-pretransform, miloslavskaya-frozen-design, Miloslavskaya-recursive-design}. However, utilizing SCL decoder with a short list size is essential to minimize decoding latency to enable URLLC in practice. Unfortunately, when using SCL decoders with short list sizes (e.g., fewer than 8), improving the code weight spectrum through pre-transformation does not necessarily lead to better decoding performance.

The SCL decoder can fail to decode the transmitted codeword under two exclusive error conditions. The first occurs when the transmitted codeword is not in the final list of the codewords identified by the SCL decoder. The second error happens when the transmitted codeword is present in the final list, but another codeword in the list is closer to the received vector. As a result, under a short list size of SCL decoder, it is essential to design a pre-transform method that can reduce the two exclusive error events simultaneously.  

In this paper, we advance the design of pre-transformed polar codes to enhance decoding performance with SCL decoders using limited list sizes. Our key innovation is to apply two different types of polar pre-transformation methods in parallel, with each method targeting the reduction of one of two distinct error event sets simultaneously. This improvement is crucial for enabling the next generation of URLLC.

\subsection{Related Works}

Since the introduction of polar codes \cite{arikan-polar}, proven to achieve asymptotic capacity over binary input discrete memoryless channels (B-DMC) with non-random construction and low-complexity successive cancellation (SC) decoding algorithms, many studies have focused on selecting information index sets for various channels and SC decoding methods \cite{arikan-polar, Tal-polar-construction, Mori-density-evolution, Trifnov-polar-construction, Wu-DEGA, schurch-partial-order, Mondelli-partial-order, beta-expansion, vangala2015comparative}. In the seminal paper \cite{arikan-polar}, Bhattacharyya parameters were used as reliability metrics, with a simple recursive relation for binary erasure channels (BEC) and Monte-Carlo simulations required for other channels. For efficient construction, a method to approximate bit-channels with large output alphabets by those with smaller alphabets was presented in \cite{Tal-polar-construction}. For additive white Gaussian noise (AWGN) channels, density evolution (DE) \cite{Mori-density-evolution} tracks log-likelihood ratios (LLR) through the SC decoding process. To reduce DE complexity, Gaussian approximations (GA) were proposed in \cite{Trifnov-polar-construction, Wu-DEGA}, approximating LLRs as Gaussian random variables to estimate their means. The information index sets from these methods depend on channel types and qualities. Additionally, channel-independent methods based on partial order have been developed in \cite{schurch-partial-order, Mondelli-partial-order, beta-expansion}.

In the finite blocklength regime, channel polarization effect is incomplete, leading to performance degradation of SC decoders due to the dominance of the worst bit-channel error probability. To overcome this problem, list decoding method was considered as an efficient decoder for the polar codes \cite{Tal-polar-SCL}. 
Unlike the SC decoder, however, finding the optimal information index set is a very challenging under the use of SCL decoder. The main difficulty arises from the fact that the DE methods developed under SC decoder \cite{arikan-polar, Tal-polar-construction, Mori-density-evolution, Trifnov-polar-construction, Wu-DEGA, schurch-partial-order, Mondelli-partial-order, beta-expansion} do not guarantee the optimally for the performance of SCL decoder. One approach to resolve this issue was to harness the path metric range of the SCL decoder to replace some frozen bits with information bits, maintaining a large path metric range \cite{Rowshan-bit-swapping}. Another relevant prior work was in  \cite{Coskun-SCL-entropy}, in which the required list size of SCL decoder was identified to achieve ML decoding performance. As subsequent work, some genetic type algorithm was proposed in \cite{miloslavskaya-frozen-design} to generate the optimized information set for SCL decoding, which showed a better decoding performance than that in \cite{Coskun-SCL-entropy}. Notwithstanding these advancements, there is still no optimal design method to determine the information set for a given code rate, blocklength, and list size.


Identifying minimum-weight (min-weight) codewords in polar codes is crucial for designing pre-transform techniques to reduce their number \cite{Dragoi-partial-order, rowshan-minweight-formation, Rowshan-minweight-pretransform, Dragoi-1.5d, zunker-enumeration-pretransform}. For example, the number of min-weight codewords was characterized by examining the automorphism group in decreasing monomial codes \cite{Dragoi-partial-order}. Further, a study demonstrates how combining rows of the polar transform matrix generates min-weight codewords and shows that carefully replacing information indices with frozen indices can reduce their number \cite{rowshan-minweight-formation}. This insight is used to explore the impact of pre-transforms, confirming their ability to effectively eliminate min-weight codewords \cite{li-pretransformed}.


Pre-transformed polar codes encompass various instances, including cyclic-redundancy-check-aided polar (CA-polar) \cite{Niu-CA-polar}, polar codes with dynamic frozen bits \cite{Trifonov-polar-dynamic-frozen}, parity-check polar (PC-polar) \cite{Wang-PCC-polar, Zhang-PC-polar-Huawei}, PAC codes \cite{arikan-pac}, row-merged polar codes \cite{Gelincik-polar-row-merging, zunker-row-merged}, and deep polar codes \cite{deep-polar-construction-A,deep-polar-gc-wkshps}. Numerous studies focus on optimizing the pre-transform matrix to reduce the number of low-weight codewords \cite{Chiu-PAC-construction, Miloslavskaya-design-short-polar, zunker-enumeration-pretransform}. Numerical results show that pre-transformation significantly enhances the weight spectrum of codes. For example, PAC codes with convolutional precoding achieve the normal approximation bound with Fano decoding for short block lengths \cite{polyanskiy-finite}. Additionally, deep polar codes with a multi-layered polar pre-transform have been shown to reach the normal approximation bound, especially with the SCL decoder featuring a large list size and backpropagation parity check (SCL-BPC). Despite these improvements, pre-transformed polar codes still require an ML-like decoder to approach the theoretical bound.


Finding an efficient pre-transform matrix and rate-profile method for a low-complexity SCL decoder remains a significant challenge. Several studies focus on optimizing the pre-transform matrix to reduce low-weight codewords \cite{Chiu-PAC-construction, Miloslavskaya-design-short-polar, Gelincik-polar-row-merging, Rowshan-minweight-pretransform, zunker-enumeration-pretransform}. Rate-profile is typically determined by i) the Reed-Muller rule for SCL decoding with large list sizes (i.e., ML-like decoding) and ii) a polarization-based rule for SC decoding error probability. However, the path metric of the decoding candidate must be jointly taken into account for SCL decoding with smaller or moderate list sizes. The most prior studies in \cite{miloslavskaya-frozen-design, Miloslavskaya-recursive-design} have addressed the complexity issue of SCL decoding with relatively large list sizes, ranging from 32 to 1024. Unfortunately, optimizing the pre-transform with rate profiling for small list sizes in SCL decoding remains an unresolved issue. In this paper, we tackle this problem by introducing a new pre-transform technique for polar codes.


\subsection{Contributions}


 Our contributions are summarized as follows:

\begin{itemize}
    \item  We present a novel pre-transform polar code called sparsely pre-transformed polar (SPP) codes. The SCL decoder can fail to decode the codeword due to two distinct error events. To address this, we propose exploiting two different types of pre-transformations based on polar kernel matrices in parallel. Each pre-transform is specialized to mitigate one of the two error event sets. Specifically:

\begin{itemize}
    \item One type of pre-transform aims to prevent the use of consecutive unreliable information bits, reducing the error event where the transmitted codeword is not in the final list of the SCL decoder.
    \item The other type of pre-transform targets reducing the number of low-weight codewords, decreasing the error event where another codeword is closer to the received signal than the transmitted codeword when both are in the final list.
\end{itemize}

\item  Our parallel pretransform structure enables us to simultaneously reduce the two distinct error events. This is a key distinction from our prior work on deep polar codes \cite{deep-polar-construction-A}, where the polar-based pretransforms were applied serially, focusing only on reducing the number of low-weight codewords. Despite this change, the proposed parallel pretransform structure retains the advantage of low-complexity encoding by leveraging the small sizes of polar kernel-based pretransforms, which results in a sparse pretransform matrix structure.

\item   We present theoretical justification for our proposed code construction. We use path metric range and the entropy of the SCL decoder's decoding path to argue that consecutive semi-polarized bits increase the decoding error probability by increasing the likelihood of the correct decoding path being deleted. Additionally, we utilize the formation of min-weight codewords to demonstrate that our rate-profile and row-merging pair selection algorithm efficiently eliminate min-weight codewords.

 \item Our simulations demonstrate that our proposed method achieves state-of-the-art block error rates (BLERs) compared to all existing finite block length codes, including 5G CRC-aided polar codes, PAC codes, and deep polar codes, across various code rates and short block lengths with SCL decoding. The performance gains are especially pronounced when employing SCL decoding with a small list size.


\end{itemize}

  





\section{Preliminaries} 
In this section, we explain some preliminaries that are relevant to this work.

\subsection{Channel Coding System}
We denote the information vector by ${\bf d} = [d_1, d_2, \ldots, d_K] \in \{0,1\}^K$, where each $d_k$ is an independent and uniformly distributed random variable over $\{0,1\}$ for $k \in [K]$. An encoder $\mathcal{E}:\{0,1\}^K\rightarrow \{0,1\}^N$ is employed to map the information vector ${\bf d}$ to a binary codeword ${\bf x} = [x_1, x_2, \ldots, x_N]$ of length $N$. The code rate $R$ is defined as the ratio of transmitted information bits to the code block length, i.e., $R = \frac{K}{N}$.

Consider a binary input discrete memoryless channel (B-DMC) represented by the mapping $W: \mathcal{X}\rightarrow \mathcal{Y}$, where $\mathcal{X}=\{0,1\}$ is the binary input alphabet and $\mathcal{Y}$ is the arbitrary output alphabet. The transmission involves sending a binary codeword ${\bf x}$ through this channel, resulting in an output sequence ${\bf y}=[y_1, y_2, \ldots, y_N]$ of length $N$.

A decoder, denoted as \(\mathcal{D}:\mathcal{Y}^N\rightarrow \mathcal{X}^K\), is employed to produce an estimate of the message bits, represented as \(\hat{\mathbf{d}}\). The primary objective of the decoder is to minimize the BLER. The BLER is defined as the probability of incorrectly estimating the transmitted message as \begin{align}
	 {\sf P}(E) =\mathbb{P}[ {\bf d}\neq {\bf \hat d}],
\end{align}
and the decoder aims to optimize its performance in achieving accurate and reliable decoding.

Given channel $W$, we can establish the channel parameters associated with the probability of error, $P(E)$. Specifically, we focus on defining two crucial parameters: the symmetric channel capacity and the Bhattacharyya parameter. The symmetric capacity of the B-DMC is defined as follows:
\begin{align}
	I(W) \defeq\! \!\!\sum_{x_n\in  \mathcal{X}}\sum_{y_n\in \mathcal{Y}}\frac{1}{2}W(y_n|x_n)\log_2\frac{W(y_n|x_n)}{\frac{1}{2}W(y_n|0)+\frac{1}{2}W(y_n|1)}. \label{eq:bit-channelcap}
\end{align}
The Bhattacharyya parameter for B-DMC is formally defined as
\begin{align}
	Z(W) \defeq \sum_{y_n\in \mathcal{Y}} \sqrt{W(y_n|0)W(y_n|1)}. \label{eq:bit-channelBH}
\end{align}

We also define the minimum distance of linear block code $\mathcal{C}$ as
\begin{align}
d^{\rm min} \defeq \min_{{\bf x} \in \mathcal{C}\backslash \{\bf 0\} } {\sf wt}({\bf x}), 
\end{align}
where ${\sf wt}({\bf x}) \defeq \Vert {\bf x} \Vert_0$ represents the number of ones in ${\bf x}$. We partition the linear block code $\mathcal{C}$ based on the number of ones in its codewords, leading to the decomposition $\mathcal{C} = \bigcup_d \mathcal{C}_d$, where $\mathcal{C}_d \defeq \{ {\bf x} \in \mathcal{C}: {\sf wt} ({\bf x}) = d \}$. Consequently, the weight spectrum of a linear block code is defined as follows:
\begin{align}
A(\mathcal{C}) \defeq \left\{(d, A_d): 0\le d \le N, ~ A_d = \vert \mathcal{C}_d \vert \right\} .
\end{align}
The two parameters, ${d}^{ \rm  min}$ and $A_{{d}^{\rm min}}$,  play a crucial role in determining the performance of a code under  ML decoding. By increasing the minimum distance and simultaneously reducing the number of codewords with minimum weights, it is possible to significantly improve the code performance. Under ML decoding, the BLER is upper bounded by the union bound as
\begin{align}
    {\sf P}^{\sf ML}(E)\leq \sum_d A_d Q\left(\sqrt{2dR\frac{E_{\rm b}}{N_{\rm 0}}}\right),
\end{align}
where $Q(u)=\int_{u}^{\infty}\frac{1}{\sqrt{2\pi}}e^{-\frac{x^2}{2}}{\rm d}x$ and $\frac{E_{\rm b}}{N_{\rm 0}}$ denotes the energy per bit to noise density ratio.

\subsection{Polar Codes}
A polar code with parameters $(N,K,\mathcal{I})$ is characterized by a polar transform matrix of size $N=2^n$ and an index set $\mathcal{I}\subseteq [N]$ where $[N]=\{1, \ldots, N\}$. Unless otherwise stated, we assume indices starting from one. The polar transform matrix of size $N=2^n$ is obtained through the $n$th Kronecker power of a binary kernel matrix ${\bf G}_2=\begin{bmatrix}
1 & 0\\
1 & 1 
\end{bmatrix}$ as
\begin{align}
	{\bf G}_N = {\bf G}_2^{\otimes n}.
\end{align}
The input vector of the encoder, denoted as ${\bf u}=[u_1,u_2,\ldots, u_N] \in \mathbb{F}_2^N$, is generated based on the given information set $\mathcal{I}$. In this process, the data vector carrying $K$ information bits, denoted as ${\bf d}$, is allocated to ${\bf u}_{\mathcal{I}}$. The remaining elements of ${\bf u}$, denoted by ${\bf u}_{\mathcal{I}^c}$, are assigned zeros. Here, $\mathcal{I}^c=[N] \backslash \mathcal{I}$ is referred to as the frozen bit set. This data assignment procedure is commonly known as rate-profiling. Finally, a polar codeword is constructed by multiplying ${\bf u}$ with ${\bf G}_N$ as 
\begin{align}
	{\bf x}^{\sf Polar}={\bf u}{\bf G}_N=\sum_{i\in \mathcal{I}} u_i {\bf g}_{N,i},
\end{align}
where ${\bf g}_{N,i}$ is the $i$th row vector of ${\bf G}_N$. 

After channel combining and splitting \cite{arikan-polar}, the $i$th bit-channel $W_N^{(i)}: \mathcal{X}\rightarrow \mathcal{Y}^N\times \mathcal{X}^{i-1}$, where $i\in [N]$, is defined as follows:
\begin{align}
   W_N^{(i)}\left({\bf y}, {\bf u}_{1:i-1} | u_i\right) = \sum_{{\bf u}_{i+1:N} \in \mathbb{F}_2^{N-i}} \frac{1}{2^{N-1}} W^N\left({\bf y}  | {\bf x}\right),
\end{align}
where $W^N\left({\bf y} | {\bf x}\right) = \prod_{i=1}^N W(y_i | x_i)$ is the  $N$ copies of B-DMCs and ${\bf u}_{a:b}=[u_a,u_{a+1},\ldots, u_{b}]$ for $a,b\in [N]$ and $a<b$. In the case of an infinite block length, the bit-channels are perfectly polarized into two states, i.e., $I\left(W_N^{(i)}\right) \rightarrow 0$ or $I\left(W_N^{(i)}\right) \rightarrow 1$ as $N\rightarrow \infty$. As a result, when $N$ is sufficiently large enough, the proper rate profiling is to select the indices having the capacity of one, i.e., 
\begin{align}
	\mathcal{I}=\left\{ i\in [N] : I\left(W_N^{(i)} \right) =1-\epsilon \right\},
\end{align}
for small $\epsilon>0$. This rate profiling is sufficient to achieve the capacity under simple SC decoding \cite{arikan-polar}.  For a short blocklength regime under SCL decoding, however, it remains open how to optimally choose the information set for polar codes.

\subsection{Pre-Transformed Polar Codes}
A $(N,K,\mathcal{I},{\bf T})$ pre-transformed polar code comprises a binary upper-triangular pre-transformation matrix ${\bf T}\in \mathbb{F}_2^{N\times N}$ and an information set $\mathcal{I}$ used for rate profiling. During encoding, the information vector ${\bf d}\in \mathbb{F}^K$, containing $K$ information bits, is incorporated into the input vector ${\bf v}\in \mathbb{F}^N$ of the pre-transformation matrix. After setting ${\bf v}_{\mathcal{I}}={\bf d}$ and ${\bf v}_{\mathcal{I}^c}={\bf 0}$, the codeword ${\bf x}$ is generated as 
\begin{align}
	{\bf x}={\bf v}{\bf T}{\bf G}_N,
\end{align}
where ${\bf G}_N = {\bf G}_2^{\otimes n}$ is polar transform matrix of size $N=2^n$ and ${\bf G}_2=\begin{bmatrix}
1 & 0\\
1 & 1 
\end{bmatrix}$.

Selecting an appropriate precoding matrix ${\bf T}\in \mathbb{F}_2^{N\times N}$ along with rate profiling set $\mathcal{I}$ gives rise to a significant challenge when constructing the pre-transformed polar codes. The joint optimization of ${\bf T}$ and $\mathcal{I}$ to maximize the weight spectrum of a code necessitates a highly complex optimization process, even for short blocklength scenarios. Optimizing becomes more complex when factoring in the restricted decoding capabilities, specifically, a SCL decoder with a small list size.

\subsection{SCL Decoding for Pretransformed Polar Codes}

\begin{figure}[t]
\centering
\includegraphics[width=1\columnwidth]{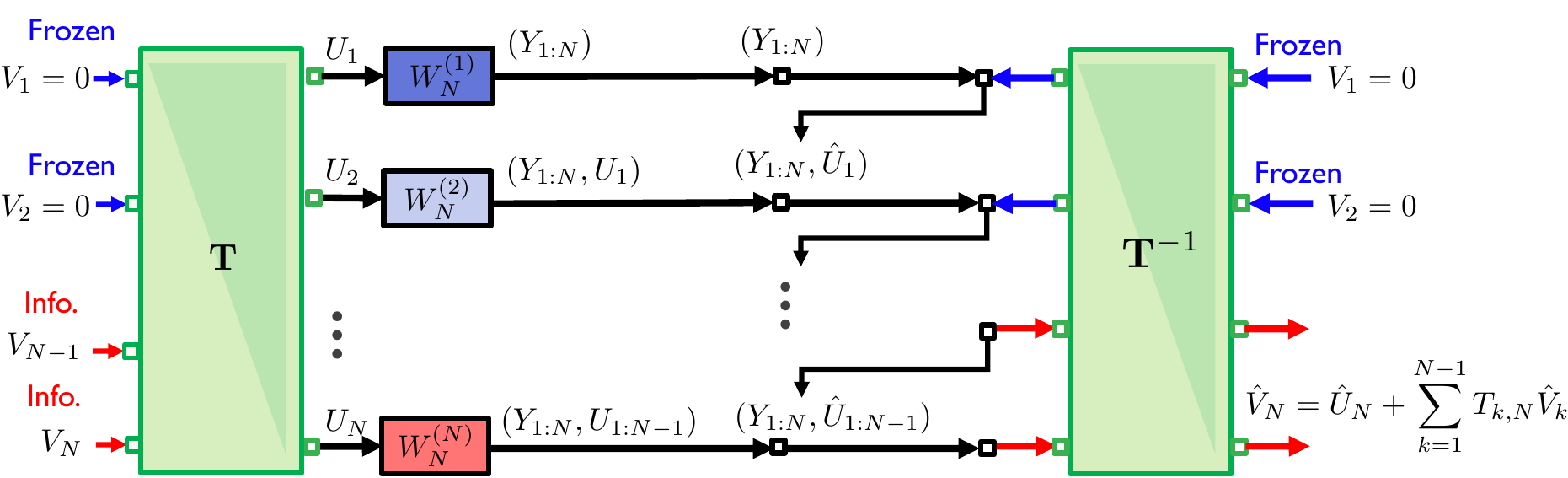}
\caption{Decoding of pre-transformed polar codes}
\label{fig:general-decoder}
\end{figure}

We explain SCL decoding of general pre-transformed polar codes shown in Fig.~\ref{fig:general-decoder}. Recall that input vector ${\bf v}$ is pre-transformed into ${\bf u}$ by upper-triangular matrix ${\bf T}$ (i.e., ${\bf u} = {\bf v}{\bf T})$, followed by polar transform ${\bf G}_N$ to generate codeword ${\bf x}={\bf u}{\bf G}_N$. The codeword ${\bf x}$ is mapped into corrupted codeword ${\bf y}$ by channel $W({\bf y}\vert {\bf x})$. The decoding proceeds in reverse order. Given noisy codeword ${\bf y}$, the decoder sequentially estimate $\hat{u}_i$ from $i=1$ to $N$. Then, the inverse pre-transform is applied to retrieve $\hat{\bf v}_{1:N} = \hat{\bf u}_{1:N} {\bf T}^{-1}$. 

When decoding $\hat{u}_i$, the decoder utilizes the polarized bit channel $W_N^{(i)}({\bf y}, {\bf u}_{1:i-1} \allowbreak \vert u_i)$. As to previous bits ${\bf u}_{1:i-1}$, the decoder presumes the estimated bits $\hat{\bf u}_{1:i-1}$ are true. 
To increase the probability of correctness of this assumption, the SCL decoder keeps $S$ hypotheses of previous estimate bits. The $i$th bit of the $s$th hypothesis is denoted as $\hat{u}_{i}[s]$. For each decoding path $s$ and bit index $i$, the decoder duplicates the current hypothesis $\hat{\bf u}_{1:i-1}[s]$ into $(\hat{\bf u}_{1:i-1}[s], 0)$ and $(\hat{\bf u}_{1:i-1}[s], 1)$. After that, the decoder prunes the decoding paths, which violate the frozen bit condition, i.e., $\hat{\bf v}_{i\in\mathcal{F}}[s] = 0$. In addition, if the number of decoding paths exceeds the predetermined threshold $S$, the most $S$ probable paths in terms of $W_N^{(i)}({\bf y}, \hat{\bf u}_{1:i-1}[s] \mid u_i = \hat{u}_i[s])$ are maintained and remaining paths are eliminated. 
For the recursive computation of $W_N^{(i)}({\bf y}, \hat{\bf u}_{1:i-1}[s] \mid u_i = \hat{u}_i[s])$, we refer the reader to \cite[Theorem 1]{Balatsoukas-SCL-log}.

In the decoding process, we need to check frozen bit condition in $v$-domain, while the decoding proceeds in $u$-domain. Thanks to upper-triangular structure of pre-transform matrix, we can retrieve ${\bf v}_{1:i}$ from ${\bf u}_{1:i}$ by
\begin{align}
    {\bf v}_{1:i} = {\bf u}_{1:i} \left({\bf T}_{1:i}\right)^{-1},
\end{align}
where ${\bf T}_{1:i}$ is upper-left sub-matrix with size $i\times i$. The same process can be performed in recursively:
\begin{align}
    v_i = u_i + \sum_{k=1}^{i-1} T_{k, i} v_k.
\end{align}

 \subsection{Notations}
We provide some definitions. We denote scalar quantity as lowercase letter $x$ and corresponding random variable as uppercase letter $X$. We denote vector as bold lowercase letter ${\bf x}$ and corresponding random vector as bold uppercase letter ${\bf X}$. Given a vector ${\bf x}$, we generally use one-based indexing, i.e., ${\bf x} = [x_1, \ldots, x_N]$. Exclusively, zero-based indexing is used in Section~\ref{sec:principle}. Given a set $\cA$ and vector ${\bf x}$, for example, if $\cA=\{1,3\}$, we use the notation ${\bf x}_{\cA}=[x_1, x_3]$. Let $[a,b]$ be the set denoting $\{a, a+1, \ldots, b\}$. Given index $i \in [0, 2^n-1]$, the binary representation of $i=\sum_{k=0}^{n-1} i_k 2^k$ is defined as ${\sf bin}(i)\defeq[i_{n-1}, i_{n-2}, \ldots, i_1, i_0]$ and support of ${\sf bin}(i)$ as $\cS_i = {\sf supp}({\sf bin}(i))\defeq \{k \in [0,n-1]: i_k = 1\}$. 
Let $\cI \subseteq [0, N-1]$ be the set of information indices satisfying the partial order property \cite{Dragoi-partial-order, schurch-partial-order, Mondelli-partial-order, beta-expansion} and $\cF = \cI^c = [0, N-1]\backslash \cI$.
Define the set $\cK_i \defeq \{ j \in \cI \backslash [0, i]: |\cS_j \backslash \cS_i| = 1\}$. It is equivalent to $\cK_i = \{ j \in [i+1, N-1]: {\sf wt}({\bf g}_{N,j}) \ge {\sf wt} ({\bf g}_{N,i} + {\bf g}_{N,j}) = {\sf wt}({\bf g}_{N,i})\}$ according to \cite[Lemma 2]{rowshan-minweight-formation}.

\section{ SCL Decoding Error Analysis}\label{sec:principle}

In this section, we provide an error analysis for the SCL decoder. Through this analysis, we demonstrate how two different types of error events determine the SCL decoding performance depending on the list size. We then present methods to reduce each of these error events separately, using rate profiling and pre-transform techniques. These techniques are jointly used to construct our SPP codes. 



\subsection{Two Error Events in SCL Decoding}
The decoding error event \(\mathcal{E}\) can be decomposed into two mutually exclusive event sets: \(\mathcal{E}_1\) and \(\mathcal{E}_2\). Here, \(\mathcal{E}_1\) represents the error event where the transmitted codeword does not appear in the final list of the SCL decoder. On the other hand, \(\mathcal{E}_2\) denotes the case where another codeword in the list is closer to the received signal than the transmitted codeword. Therefore, the decoding error probability under SCL decoding with a list size of \(S\) can be expressed as:
\begin{align}
    P_{\sf SCL} (\cE; S) = P_{\sf SCL} (\cE_1; S) + P_{\sf SCL} (\cE_2; S).
\end{align}
The error event $\cE_1$ depends on the selection of information index set $\cI$, i.e., rate-profile. The event $\cE_2$ depends on minimum distance and the number of minimum weight (min-weight) codewords, which is a function of rate-profile $\cI$ and pre-transform matrix ${\bf T}$. 

If $S=1$, then SCL decoder reduces to SC decoder and decoding error probability of SC decoder $P_{\sf SC}(\cE)$ is given by
\begin{align}
    P_{\sf SC}(\cE) &= P_{\sf SCL}(\cE; S=1)  \\
    &= P_{\sf SCL}(\cE_1; S=1) \\
    &= 1 - \prod_{i=0}^{N-1} \left(1 -  P\left(\cE; W_N^{(i)} \right) \right),
\end{align}
where $P\left(\cE; W_N^{(i)} \right)$ is error probability of channel $W_N^{(i)}$, given by
\begin{align}
    & P\left(\cE; W_N^{(i)} \right) \nonumber \\
    &= \mathbb{P}\left( W_N^{(i)} ({\bf y}, {\bf u}_{0:i-1} \vert 0) < W_N^{(i)} ({\bf y}, {\bf u}_{0:i-1} \vert 1) \mid {\bf u} = {\bf 0} \right) \nonumber \\
    & \quad + \frac{1}{2} \mathbb{P}\left( W_N^{(i)} ({\bf y}, {\bf u}_{0:i-1} \vert 0) = W_N^{(i)} ({\bf y}, {\bf u}_{0:i-1} \vert 1) \mid {\bf u} = {\bf 0} \right).
\end{align}
In this case, decoding error event is dominated by information index set. For example, if a channel $W$ is BEC, then $P\left(\cE; W_N^{(i)}\right)$ is directly a function of bit-channel capacity $I\left(W_N^{(i)}\right)$. Therefore, the design of pre-transformed polar codes becomes finding the $K$ most reliable bit-channels, which is well studied in \cite{arikan-polar, Tal-polar-construction, Mori-density-evolution, Trifnov-polar-construction, Wu-DEGA, schurch-partial-order, Mondelli-partial-order, beta-expansion}.

On the contrary, if $S=\infty$, then SCL decoder becomes ML decoder and decoding error probability is given by 
\begin{align}
    P_{\sf SCL}(\cE; S=\infty) &= P_{\sf SCL}(\cE_2; S=\infty) \\
    &\lessapprox A_{d_{\rm min}} Q\left(\sqrt{2d_{\rm min} R \frac{E_{\rm b}}{N_0}}\right).
\end{align}
 In this scenario, the decoding error event is primarily influenced by the minimum distance \(d_{\text{min}}\) and the number of min-weight codewords \(A_{d_{\text{min}}}\). Since both \(d_{\text{min}}\) and \(A_{d_{\text{min}}}\) depend on the rate-profile \(\mathcal{I}\) and the pre-transform matrix \(\mathbf{T}\), jointly optimizing both \(\mathcal{I}\) and \(\mathbf{T}\) is very challenging to construct the pre-transformed polar codes achieving high BLER performance under SCL decoding with a small list size. Most prior work has focused on finding pre-transform matrices that minimize \(A_{d_{\text{min}}}\) with a fixed \(\mathcal{I}\) (i.e., RM profiling)   \cite{Chiu-PAC-construction, Miloslavskaya-design-short-polar, zunker-enumeration-pretransform}.


\subsection{Rate Profiling Method to Reduce $P_{\sf SCL} (\cE_1; S)$}\label{sec:consecutive info bit}

We explain how to design information sets to diminish \(P_{\sf SCL}(\mathcal{E}_1; S)\). Our principle for reducing the error event, where the correct codeword does not exist in the list, is to avoid using less reliable information bits consecutively. To elucidate this principle, we adopt the decoder entropy analysis tool introduced in \cite[Theorem 1]{Coskun-SCL-entropy}, which was originally proposed to determine the required list size for the SCL decoder to achieve ML decoding performance.

Let $\cI^{(m)} \defeq \cI \cap [0, m-1]$ and $\cF^{(m)} \defeq \cF \cap [0, m-1]$ be the sets containing information and frozen indices within the first $m$ input bits, respectively. 
The uncertainty of decoding paths when decoding $u_m$ is characterized by the entropy of ${\bf U}_{\cI^{(m)}}$ given ${\bf Y} = {\bf y}$ and ${\bf U}_{\cF^{(m)}}$ as
\begin{equation}
    d_m({\bf y}) \defeq H \left( {\bf U}_{\cI^{(m)}} \vert {\bf Y}={\bf y}, {\bf U}_{\cF^{(m)}} \right).
\end{equation}
From this definition, we can define the corresponding random variable $D_m$, which takes  the value $d_m({\bf y})$ when ${\bf Y} = {\bf y}$ according to the conditional distribution. To measure the uncertainty on $d_m({\bf Y})$, we define the corresponding conditional entropy as
\begin{equation}
    \bar{D}_m \defeq \mathbb{E} [d_m({\bf Y})] = H\left( {\bf U}_{\cI^{(m)}} \vert {\bf Y}, {\bf U}_{\cF^{(m)}} \right).
\end{equation}
This conditional entropy, \(\bar{D}_m\), can be interpreted as decoding uncertainty. Consequently, it plays an important role in determining the required list size to achieve ML decoding performance under SCL decoding \cite[Theorem 1]{Coskun-SCL-entropy}.
 
To quantify the sole impact of decoding uncertainty introduced by the $m$th bit $u_m$, we define the incremental entropy by decoding $u_m$ by the difference between $\bar{D}_m$ and $\bar{D}_{m-1}$ as follows:
\begin{equation}
    \Delta_m \defeq \bar{D}_m - \bar{D}_{m-1}.  \label{eq:Entropy_Dec_m}
\end{equation}
The following lemma demonstrates that decoding uncertainty can increase or decrease depending on whether \( u_m \) is an information bit or a frozen bit. \vspace{0.2cm}
\begin{lemma}\label{lem1} The incremental entropy by decoding $u_m$ is 
\begin{align}
     \begin{cases}
		\Delta_m	\geq 0, & \text{if $m\in \mathcal{I}$ }\\
          \Delta_m  \leq 0, & \text{otherwise}.
		 \end{cases}
\end{align}

\end{lemma}

\begin{IEEEproof}
Although the proof is available in \cite{Coskun-SCL-entropy}, we include it here for completeness. Suppose $m \in \cI$. Then,
\begin{align}
    & \Delta_m = H\left( {\bf U}_{\cI^{(m)}} \vert {\bf Y}, {\bf U}_{\cF^{(m)}} \right) - H\left( {\bf U}_{\cI^{(m-1)}} \vert {\bf Y}, {\bf U}_{\cF^{(m-1)}} \right) \nonumber \\
    &= H\left( {\bf U}_{\cI^{(m)}} \vert {\bf Y}, {\bf U}_{\cF^{(m-1)}} \right) - H\left( {\bf U}_{\cI^{(m-1)}} \vert {\bf Y}, {\bf U}_{\cF^{(m-1)}} \right) \nonumber \\
    &= H\left( {\bf U}_{\cI^{(m)}}, {\bf U}_{\cF^{(m-1)}} \vert {\bf Y} \right) - H\left( {\bf U}_{\cI^{(m-1)}}, {\bf U}_{\cF^{(m-1)}} \vert {\bf Y} \right) \nonumber \\
    &= H({\bf U}_{0:m-1} \vert {\bf Y}) + H(U_m \vert {\bf Y}, {\bf U}_{0:m-1}) - H({\bf U}_{0:m-1} \vert {\bf Y}) \nonumber \\
    &= H(U_m \vert {\bf Y}, {\bf U}_{0:m-1}) \geq 0,\label{eqn:info-bit-entropy-increase}
\end{align}
where the last inequality follows from the fact that entropy is always greater than or equal to zero. Conversely, if $m \in \cF$, we can compute the incremental entropy as
\begin{align}
    & \Delta_m = H({\bf U}_{\cI^{(m)}} \vert {\bf Y}, {\bf U}_{\cF^{(m)}} ) - H({\bf U}_{\cI^{(m-1)}} \vert {\bf Y}, {\bf U}_{\cF^{(m-1)}}) \nonumber \\
    &= H({\bf U}_{\cI^{(m-1)}} \vert {\bf Y}, {\bf U}_{\cF^{(m-1)}}, U_m) - H({\bf U}_{\cI^{(m-1)}} \vert {\bf Y}, {\bf U}_{\cF^{(m-1)}} ) \nonumber \\
    &= -I(U_m; {\bf U}_{\cI^{(m-1)}} \vert {\bf Y}, {\bf U}_{\cF^{(m-1)}}) \leq 0. \label{eqn:frozen-bit-entropy-decrease}
\end{align}
\end{IEEEproof} 
\vspace{0.2cm}


From Lemma \ref{lem1}, if the information bits are chosen consecutively (e.g., \( m, m+1 \in \mathcal{I} \)), the decoder entropy continuously increases. The amount of this increment depends on the bit-channel capacities of channels \( W_N^{(m)}({\bf y}, {\bf u}_{0:{m-1}} | u_m) \) and \( W_N^{(m+1)}({\bf y}, {\bf u}_{0:m} | u_{m+1}) \). When \( u_m \) and \( u_{m+1} \) are sent over sufficiently polarized bit-channels with \( I\left(W_N^{(m)}\right) = I\left(W_N^{(m+1)}\right) \approx 1 \), the decoder entropy does not increase, i.e., \( H(U_m | {\bf Y}, {\bf U}_{0:m-1}) = H(U_{m+1} | {\bf Y}, {\bf U}_{0:m}) \approx 0 \). As a result, the SC decoder is sufficient, provided that all information bits are transmitted over sufficiently polarized bit-channels. However, in the case of finite blocklengths, such channel polarization does not occur, and some information bits must be sent over less reliable bit-channels. When these bits are sent consecutively over less reliable bit-channels, the decoder entropy keeps increasing. In other words, the decoder requires a larger list size to maintain the transmitted codeword in the list.

When the list size is small, carefully designing information sets is crucial to keeping the transmitted codeword in the list. As a result, to reduce \( P_{\sf SCL} (\cE_1; S) \), inserting frozen bits between information bits is an effective method since it prevents consecutive information bits. By inserting frozen bits between information bits, it is possible to reduce decoder entropy since \( \Delta_m < 0 \) for \( m \in \mathcal{F} \), as shown in \eqref{eqn:frozen-bit-entropy-decrease}. 
However, inserting frozen bits between consecutive information bits can increase the number of min-weight codewords, thereby increasing \( P_{\sf SCL} (\cE_2; S) \). In the next section, we will propose a rate profiling method with pre-transform, jointly considering both effects.

\subsection{Bit-Swapping Method to Reduce $P_{\sf SCL} (\cE_2; S)$}\label{sec:bit-swapping}

Given $(N,K,\cI)$ polar code, define a coset of codewords $\cC_i(\cI)$ with the coset leader ${\bf g}_{N,i}$ as
\begin{equation}
    \cC_i(\cI) \defeq \left\{ {\bf g}_{N,i} \oplus \bigoplus_{h \in \cH} {\bf g}_{N,h}: \cH \subseteq \cI \backslash [0, i] \right\}.
\end{equation}
The coset $\cC_i(\cI)$ generates min-weight codewords as follows. 
Every row ${\bf g}_{N,i}, i\in\cI$ of the polar transform matrix ${\bf G}_N$, where ${\sf wt}({\bf g}_{N,i}) = w_{\rm min}$, can form a min-weight codeword in combination with the rows in every subset $\cJ \subseteq \cK_i$ and the corresponding set $\cM(\cJ)\subseteq (\cI \cap [i+1, N-1]) \backslash \cK_i$ as
\begin{equation}
    {\sf wt} ( {\bf g}_{N,i}  \oplus \underbrace{\bigoplus_{j \in \cJ} {\bf g}_{N,j}}_\text{ core rows} \oplus \underbrace{\bigoplus_{m \in \cM(\cJ)} {\bf g}_{N,m}}_\text{ balancing rows}) = w_{\rm min}. \label{eqn:min-weight-formation}
\end{equation}
The rows in the set $\cJ$ are called {\it core} rows.
The rows in the set $\cM(\cJ)$ are called {\it balancing} rows as their inclusion brings the weight of the sum down to $w_{\rm min}$ if needed. 
The set $\cM(\cJ)$ can be constructed by the $\cM$-Construction described in \cite[Section III-A]{rowshan-minweight-formation}.

The equation \eqref{eqn:min-weight-formation} shows one way to generate min-weight codewords, which gives a lower bound of the number of min-weight codewords of $(N,K,\cI)$ polar codes, denoted by $A_{d_{\rm min}}(\cI)$. 
Since every subset $\cJ$ of $\cK_i$ corresponds to a different min-weight codeword, the total number of such codewords in every coset $\cC_i(\cI)$ is lower bounded by the total number of subsets of $\cK_i$, that is $2^{|\cK_i|}$. This lower bound is matched with the result in \cite[Propositions 6 and 7]{Dragoi-partial-order}, which counts the number of min-weight codewords by harnessing the fact that lower-triangular affine (LTA) transformation group forms automorphism of decreasing monomial codes under the properly defined group action.

The formation of a minimum-weight codeword \eqref{eqn:min-weight-formation} is complete if the information index set \(\cI\) follows a universal partial order \cite{Dragoi-partial-order, schurch-partial-order, Mondelli-partial-order, beta-expansion}. Otherwise, given a coset leader \(\mathbf{g}_{N,i}\) and some core rows \(\cJ\), it is possible for some index \( i \in \cM(\cJ) \cap \cI^c \) to exist, preventing the formation of a minimum-weight codeword. This implies that properly swapping an information bit with a frozen bit reduces the number of minimum-weight codewords, thereby diminishing \( P_{\sf SCL}(\cE_2; S) \).

To further illustrate, define the set $\cB_{w_{\rm min}}(\cI)$ and $\cB_{w_{\rm min}}(\cF)$
\begin{align}
    \cB_{w_{\rm min}}(\cI) &\defeq \{ i \in \cI : {\sf wt}({\bf g}_{N,i}) = w_{\rm min}\}, \\
    \cB_{w_{\rm min}}(\cF) &\defeq \{ i \in \cF : {\sf wt}({\bf g}_{N,i}) = w_{\rm min}\}.
\end{align}
For each $j \in \cB_{w_{\rm min}}(\cI)$, let us also define the set $\cD_j$ and $\cG_j$ as follows.
\begin{align}
    \cD_j &\defeq \{ i \in \cB_{w_{\rm min}}(\cI): j \in \cK_i\}, \\
    \cG_j &\defeq \{ i \in \cB_{w_{\rm min}}(\cF): j \in \cK_i\}.
\end{align}
Given $j \in \cB_{w_{\rm min}}(\cI)$ and $i \in \cG_j$ satisfying, 
\begin{equation}
    \left(\sum_{x \in \cD_j} 2^{|\cK_x|-1}\right) + 2^{|\cK_j|} > 2^{|\cK_i|-1}, \label{eqn:swapping-condition}
\end{equation}
according to \cite{rowshan-minweight-formation}, we have 
\begin{align}
    &A_{w_{\rm min}}(\cI^{\prime}) \le A_{w_{\rm min}}(\cI) - \eta, \\
    &\eta = \left(\left( \sum_{x \in \cD_j} 2^{|\cK_x|-1} \right) + 2^{|\cK_j|} - 2^{|\cK_i|-1} \right) > 0,
\end{align}
where $\cI^\prime = \{i\} \cup (\cI \backslash \{j\})$. This result implies that introducing less reliable min-weight rows as information indices while freezing more reliable min-weight rows, called {\it bit-swapping}, can reduce the number of min-weight codewords, thereby reducing \( P_{\sf SCL}(\cE_2; S) \).

\subsection{Bit-Swapping to Reduce both \( P_{\sf SCL}(\cE_1; S) \) and \( P_{\sf SCL}(\cE_2; S) \)}

To reduce overall error probability $P_{\sf SCL}(\cE; S)$, we consider both $P_{\sf SCL}(\cE_1; S)$ and $P_{\sf SCL}(\cE_2; S)$, simultaneously. 
Bit-swapping can reduce the number of min-weight codewords. Especially, swapping min-weight information row and frozen row satisfying \eqref{eqn:swapping-condition} reduces the number of min-weight codewords.  
Although bit-swapping reduces $P_{\sf SCL}(\cE_2; S)$, it introduces a less reliable information bit and increases $P_{\sf SCL}(\cE_1; S)$. Therefore, it is important to carefully select the bit-swapping pairs. The proposed method is to choose the bit-swapping pairs, within min-weight rows, to ensure there are no consecutive semi-polarized information bits as illustrated in Section~\ref{sec:consecutive info bit}. This approach can improve decoding performance of SCL decoder by reducing \( P_{\sf SCL}(\cE_2; S) \) with minimal loss in \( P_{\sf SCL}(\cE_1; S) \).





\subsection{Upper-Triangular Pre-Transform to Reduce \( P_{\sf SCL}(\cE_2; S) \)}

Applying upper-triangular pre-transform to input vector of polar transform can reduce the number of min-weight codewords \cite{li-pretransformed, li-pretransformed-average}. In particular, we can definitely eliminate the min-weight codewords in $\cC_i(\cI)$ by simple upper-triangular pre-transform, if several conditions are satisfied.

 \vspace{0.3cm}
\begin{theorem}\label{thm:1}
    Given $i, j \in [0, 2^n - 1]$ such that ${\sf wt}({\bf g}_{N, i}) = w_{\rm min}$ and ${\sf wt}({\bf g}_{N, j}) \ge w_{\rm min}$, suppose $j \preceq i$ with respect to universal partial order, i.e., $W_N^{(i)}$ is more reliable than $W_N^{(j)}$\cite{Dragoi-partial-order, schurch-partial-order, Mondelli-partial-order}. For ${\bf c}\in\cC_i(\cI)$, the following holds: ${\sf wt}({\bf c} + {\bf g}_{N, j}) > w_{\rm min}$.
\end{theorem}
 \vspace{0.3cm}

\begin{IEEEproof}
    Fix $i$ with ${\sf wt}({\bf g}_{N,i}) = w_{\rm min}$. 
    According to \cite[Theorem 1]{rowshan-minweight-formation}, the combination of rows in \eqref{eqn:min-weight-formation} can generate some min-weight codewords belonging to $\cC_i(\cI)$. Furthermore, all the min-weight codewords in $\cC_i(\cI)$ must satisfy the combination of rows in \eqref{eqn:min-weight-formation} because the number of possible combination achieves the upper bound of the number of min-weight codeword proved in \cite[Proposition 6]{Dragoi-partial-order} and \cite[Proposition 2]{rowshan-minweight-formation}. 
    In particular, for any $k \in \cJ \cup \cM(\cJ)$, we have $i \preceq k$ by \cite[Lemma 6]{rowshan-minweight-formation}. 
    Therefore, ${\sf wt}({\bf c} + {\bf g}_{N, j}) > w_{\rm min}$.
\end{IEEEproof}

\vspace{0.2cm}

Theorem~\ref{thm:1} provides guidance on the design of pre-transform matrix toward reducing $P_{\sf SCL}(\cE_2; S)$. 
For each min-weight information row $i \in \cB_{w_{\rm min}}(\cI)$ and subsequent frozen row $j\in\cF \cap [i+1, N-1]$ such that ${\sf wt}({\bf g}_{N, j}) \ge w_{\rm min}$, applying pre-transform ${\bf G}_2^{\top}$ to indices $(i,j)$ prevent the formation of min-weight codewords in the coset $\cC_i(\cI)$.

\section{SPP Codes}\label{sec:code}
In this section, we introduce SPP codes, which improve upon deep polar codes \cite{deep-polar-construction-A, deep-polar-gc-wkshps} under SCL decoder with small list sizes. The deep polar codes employ a serial multi-layered polar pre-transform with rate less than one. The effect of these pre-transforms can be understood by a combination of bit-swapping and pre-transform, both of which eliminate min-weight codewords. Although deep polar codes significantly decrease $P_{\sf SCL}(\cE_2; S)$, they suffer a significant loss in $P_{\sf SCL}(\cE_1; S)$, leading to performance degradation when $S$ is small. 

The SPP codes use multiple polar pre-transform matrices in parallel, each of which belongs to a {\it Type-\Rom{1}} or {\it Type-\Rom{2}} pre-transform matrix. 
The {\it Type-\Rom{1}} pre-transform uses the transpose matrix of the polar transform kernel and aims to improve the overall error probability $P_{\sf SCL}(\cE_1; S) + P_{\sf SCL}(\cE_2; S)$. It accommodates a slight loss in $P_{\sf SCL}(\cE_1; S)$ by adopting pre-transform with a rate less than one. Despite this loss, it has the potential to significantly improve $P_{\sf SCL}(\cE_2; S)$  compared to scenarios where no loss in $P_{\sf SCL}(\cE_1; S)$ is permitted. 
The {\it Type-\Rom{2}} pre-transform uses the row-merging operation and intends to improve $P_{\sf SCL}(\cE_2; S)$ without any loss in $P_{\sf SCL}(\cE_1; S)$.

We first present the encoding process of SPP codes. Next, we present rate profile algorithm to determine connection indices for which the {\it Type-\Rom{1}} pre-transforms are applied. Lastly, we present a greedy algorithm to find row-merging pair to apply the {\it Type-\Rom{2}} pre-transforms.

\begin{figure}[t]
\centering
\includegraphics[width=1\columnwidth]{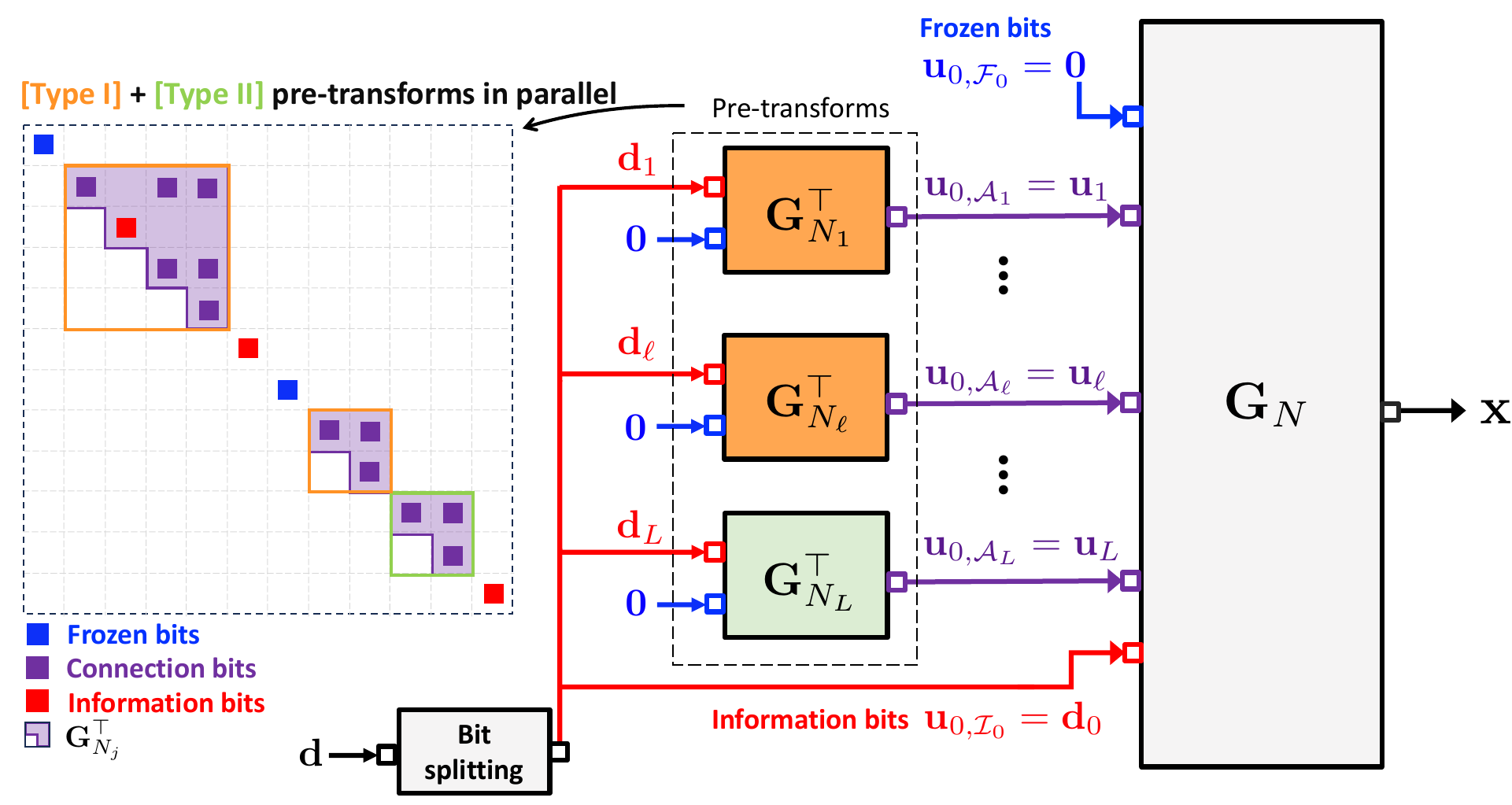}
\caption{Illustration of Encoding Structure of SPP codes. {\it Type-\Rom{1}} pre-transforms are marked in orange, and {\it Type-\Rom{2}} pre-transforms are marked in green.}
\label{fig:encoder}
\end{figure}

\subsection{Encoding}

A $(N,K, \{\mathcal{I}_{\ell}\}_{\ell=0}^{L}, \{\mathcal{A}_{\ell}\}_{\ell=1}^{L}, \{ {\bf T}_{\ell}\}_{\ell=1}^{L})$ SPP code is defined with the following parameters:
\begin{itemize}
    \item i) $L$ pre-transfom matrices ${\bf T}_{\ell} \in \mathbb{F}_2^{N_{\ell}\times N_{\ell}}$;
    \item ii) $L+1$ information sets $\{\mathcal{I}_0, \mathcal{I}_1,\ldots, \mathcal{I}_{L} \}$, and 
       \item iii) $L$ connection sets $\{\mathcal{A}_1, \mathcal{A}_2,\ldots, \mathcal{A}_{L} \}$.
\end{itemize}
The encoding process of the SPP code involves three steps: i) splitting message bits, ii) applying multiple polar pre-transforms, and iii) applying a polar transform. 

{\bf Information bit splitting and mapping:} The information vector ${\bf d}\in\mathbb{F}_{2}^{K}$ is divided into $L+1$ sub-vectors ${\bf d}_{\ell}\in \mathbb{F}_2^{K_{\ell}}$, where $K_{\ell}=|\mathcal{I}_{\ell}|$ is the number of bits allocated to ${\bf d}_{\ell}$ and $\sum_{\ell=0}^{L}K_{\ell}=K$. We denote the input vector of the $\ell$th pre-transform by ${\bf v}_{\ell}=[v_{\ell,1}, v_{\ell,2},\ldots, v_{\ell,N_{\ell}}] \in \mathbb{F}_2^{N_{\ell}}$ for $\ell\in [L]$. The index set $[N_{\ell}]$ for $\ell\in \{1,2,\ldots, L\}$ is partitioned into two non-overlapping index sets as   
\begin{align}
    [N_{\ell}] = \mathcal{F}_{\ell}\cup\mathcal{I}_{\ell}.
\end{align}
For the zeroth layer, the index set for the polar transform is divided into three non-overlapping sets as
\begin{align}
    [N] =  \mathcal{F}_{0}\cup \mathcal{I}_0\cup \mathcal{A}_0,
\end{align}
where $\mathcal{A}_0= \bigcup_{\ell=1}^L\mathcal{A}_{\ell}$ and $\mathcal{A}_{i}\cap \mathcal{A}_j =\phi$ for $i\neq j$. Each information sub-vector ${\bf d}_{\ell}$ for $\ell\in \{1,\ldots,L\}$ is assigned to the elements $v_{\ell,i}$ for $i\in \mathcal{I}_{\ell}$, i.e.,
\begin{align}
    {\bf v}_{\ell, \mathcal{I}_{\ell}}={\bf d}_{\ell}, \label{eq:infobits}
\end{align}
and the zero bits are allocated to the frozen bits in $\mathcal{F}_{\ell}$ as 
\begin{align}
    {\bf v}_{\ell, \mathcal{F}_{\ell}}={\bf 0}. \label{eq:frozenbits}
\end{align}

{\bf Parallel and local polar pre-transform:} From the information and frozen bits assignment in \eqref{eq:infobits} and \eqref{eq:frozenbits}, the input vector of the $\ell$-layer pre-transform is ${\bf v}_{\ell}=[ {\bf v}_{\ell, \mathcal{I}_{\ell}}, {\bf v}_{\ell, \mathcal{F}_{\ell}}]$. Note that we omit the index permutation for notational ease. The output vector of the $\ell$th pre-transform ${\bf u}_{\ell}\in \mathbb{F}_2^{N_{\ell}}$ is given by
\begin{align}
    {\bf u}_{\ell} ={\bf v}_{\ell}{\bf G}_{N_{\ell}}^{\top}, \label{eq:PToutput}
\end{align}
where ${\bf G}_{N_{\ell}}$ is polar transform matrix with size $N_\ell$ and $\ell\in \{1,2,\ldots, L\}$. 
For both {\it Type-\Rom{1}} and {\it Type-\Rom{2}} pre-transforms, we adopt ${\bf G}_{N_{\ell}}^{\top}$.

{\bf Global polar transform:}
Using the pre-transformed output vectors ${\bf u}_{0, \cA_{\ell}}={\bf u}_{\ell}$ for $\ell\in [L]$ in \eqref{eq:PToutput} and the information sub-vector ${\bf d}_0$, the encoder generates the input vector for the polar transform
\begin{align}
    {\bf u}_0 &= \left[{\bf u}_{0, \cF_0}, {\bf u}_{0, \mathcal{A}_1}, \ldots, {\bf u}_{0, \mathcal{A}_L}, {\bf u}_{0, \cI_0}\right] \\
    &= \left[{\bf 0}, {\bf u}_1, \ldots, {\bf u}_L, {\bf d}_0 \right].
\end{align}
Finally, applying the polar transform for the base layer, a sparesly pre-transformed codeword is constructed as
\begin{align}
    {\bf x} = {\bf u}_0 {\bf G}_N.
\end{align}
Our encoding method can be represented using a sparse pretransform matrix ${\bf T}\in \mathbb{F}_2^{N\times N}$ with a block diagonal structure as
\begin{align}
    &\left[ {\bf 0}, {\bf v}_1, \ldots, {\bf v}_L, {\bf d}_0 \right]  \underbrace{ \begin{bmatrix} 
   {\bf I} & {\bf 0}& {\bf 0} & {\bf 0} & {\bf 0} & {\bf 0} \\
   {\bf 0} & {\bf G}_1^{\top} & {\bf 0} & {\bf 0} & {\bf 0}  & {\bf 0}\\
   {\bf 0} &{\bf 0} & {\bf 0}  & \ddots & {\bf 0} & {\bf 0} \\
   {\bf 0} &{\bf 0} & {\bf 0}  & \ddots & {\bf G}_L^{\top} & {\bf 0} \\
    {\bf 0} &{\bf 0} & {\bf 0}  & \cdots & {\bf 0} & {\bf I} \\
   \end{bmatrix}}_{{\bf T}} \nonumber\\
   &=\left[{\bf u}_{0, \cF_0}, {\bf u}_{0, \mathcal{A}_1}, \ldots, {\bf u}_{0, \mathcal{A}_L}, {\bf u}_{0, \cI_0}\right].
\end{align}
The sub-block pre-transform matrices ${\bf G}_{\ell}^{\top}$ exhibit an upper triangular structure, as does ${\bf T}$. From \cite{li-pretransformed}, our pre-transform guarantees that the minimum distance of the code does not decrease after the transformation.

\begin{algorithm}
\caption{Encoding}
\KwData{$N$, $L$, $\{ \mathcal{I}_{\ell} \}_{\ell=0}^{L}$, $\{ \mathcal{A}_{\ell} \}_{\ell=1}^{L}$, ${\bf d}$.}
\KwResult{${\bf x}$.}

\smallskip
\emph{/* Bit splitting and mapping */}\;

\For{$\ell = 0, 1, \ldots, L$}{
$K_{\ell} \gets \vert \mathcal{I}_{\ell} \vert$\;
${\bf d}_{\ell} \gets {\bf d}_{1:K_{\ell}}$\;
${\bf d} \gets {\bf d}_{(K_{\ell}+1):{\rm end}}$\;
}

\medskip

\emph{/* Pre-transform */}\;
${\bf u} \gets {\bf 0}_{N}$\;
\For{$\ell = 1, 2, \ldots, L$}{
$N_{\ell} \gets \vert \mathcal{A}_{\ell} \vert$\;
${\bf v} \gets {\bf 0}_{N_{\ell}}$\;
${\bf v}_{\mathcal{I}_{\ell}} \gets {\bf d}_{\ell}$\;
${\bf u}_{\mathcal{A}_{\ell}} \gets {\bf v} {\bf G}_{N_{\ell}}^{\top}$\;
}

\medskip
\emph{/* Polar transform */}\;
${\bf u}_{\mathcal{I}_0} \gets {\bf d}_0$\;
${\bf x} \gets {\bf u} {\bf G}_N$;

{\bf Return} ${\bf x}$.

\end{algorithm}

\subsection{Type-\Rom{1} Pre-Transform and Rate Profile} \label{sec:RP}

From Section~\ref{sec:principle}, we observe that replacing min-weight information index with min-weight frozen index can remove min-weight codewords. Meanwhile, non-consecutive unreliable information bits are important to lesson the influence of less reliable information bits. The {\it Type-\Rom{1}} pre-transforms implements bit-swapping, pre-transforming, and limiting consecutive semi-polarized bits simultaneously by adopting pre-transform with rate less than one.
To efficiently reduces $P_{\sf SCL}(\cE_2; S)$ with minimal backoff of $P_{\sf SCL}(\cE_1; S)$, it is important to select connection indices set into which the output of {\it Type-\Rom{1}} pre-transforms are fed.




Suppose an SPP code with $L$ pre-transform matrices, comprised of $L_1$ {\it Type-\Rom{1}} pre-transform matrices and $L_2 = L-L_1$ {\it Type-\Rom{2}} pre-transform matrices. 
We explain how to select information set for the base layer $\mathcal{I}_{0}$ and connection sets $\mathcal{A}_{\ell}$ for $\ell\in [L_1]$. For brevity, we introduce some notations. Denote the vector ${\bf u}_{a:b}=[u_a,u_{a+b},\ldots, u_b]$, where $a<b$, and $a, b \in [N]$. Define the synthesized channel 
\begin{align}
    W^{(i)}_N({\bf y},{\bf u}_{1:i-1}|u_i)=\sum_{{\bf u}_{i+1:N}\in \mathbb{F}_2^{N-i}} \frac{1}{2^{N-1}} W^N({\bf y}|{\bf x}),
\end{align}
where $W^N({\bf y}|{\bf x})$ represents the $N$ copies of B-DMCs $W(y | x)$.
Let $n_p = \sum_{\ell=1}^{L_1} N_{\ell}$ denote the number of bits required by pre-transform and introduce the ordered index set for $i_j\in[N]$ as 
\begin{align}
    \mathcal{R} = \{ i_j \in [N] : i_1, \ldots, i_{K_0 + n_p} \}
\end{align}
where $I\left(W_N^{(i_1)}\right) \ge I\left(W_N^{(i_2)}\right) \ge \cdots \ge I\left(W_N^{(i_{K_0+n_p})}\right)$
and $I\left(W_N^{(i)}\right) > I\left( W_N^{(j)} \right)$ for any $i\in\mathcal{R}$ and $j\in [N]\backslash\mathcal{R}$.

We commence with partitioning the ordered index set $\mathcal{R}$ according to row weight of polar transform matrix as 
\begin{align}
    \cB_{w}(\cR) &= \{i \in \mathcal{R}: {\sf wt}({\bf g}_{N,i}) = w\} \\
    &= \{i^w_1, i^w_2, \ldots i^w_{|\cB_{w}(\cR)|} \},
\end{align}
where $i_1^w$ is the least reliable and $i^w_{|\cB_{w}(\cR)|}$ is the most reliable index in $\cB_{w}(\cR)$. Let $w_{\rm min} = \min_{i \in \cR} {\sf wt}({\bf g}_{N, i})$. Using $\cB_{{w}_{\rm min}}(\cR)$, we generate the auxiliary set $\tilde{\cR}$ with the size of $n_p$, which is filled up with indices belonging to $\cB_{w_{\rm min}}(\cR)$. In the process, the least reliable bit comes first. If the number of elements of $\tilde{\mathcal{R}}$ is less than $n_p$, we repeat the process using the next larger row weight index set $\cB_{2w_{\rm min}}(\cR)$ and so on. With the auxiliary set $\tilde{\mathcal{R}}$, we generate information set $\mathcal{I}_0 = \mathcal{R} \backslash \tilde{\mathcal{R}}$.
Subsequently, we reorder the elements within the set $\tilde{\mathcal{R}} = \left\{\tilde{i}_1, \ldots, \tilde{i}_{n_p}\right\}$ in a naturally ascending sequence with ${\tilde i}_j <{\tilde i}_u $ for $j<u$. Using this rearranged index set, the $\ell$th layer connection set is chosen as $    \mathcal{A}_{\ell}=\left\{ {\tilde i}_{\sum_{j=1}^{\ell-1}N_{j}+1},\ldots, {\tilde i}_{\sum_{j=1}^{\ell}N_{j}}\right\} $
for $\ell\in [L_1]$. In addition, the information set $\mathcal{I}_{\ell}$ is constructed using RM-profiling such that
\begin{align}
    \mathcal{I}_{\ell}=\{i \in [N_{\ell}]: i_1,\ldots, i_{K_{\ell}} \}
\end{align}
where
\begin{align}
    {\sf wt}\left({\bf g}_{i_1,N_{\ell}}^{\top}\right)\geq {\sf wt}\left({\bf g}_{i_2,N_{\ell}}^{\top}\right)\geq \cdots \geq  {\sf wt}\left({\bf g}_{i_{K_{\ell}},N_{\ell}}^{\top}\right).
\end{align}
We summarize our rate profile in Algorithm~\ref{Alg:RP}.

The resulting connection set consists of min-weight rows and semi-polarized rows. Because each {\it Type-\Rom{1}} pre-transform has frozen bit due to $K_{\ell} / N_{\ell} < 1$ and is applied to each connection index set $\cA_{\ell}$, the number of consecutive semi-polarized bits are limited by $\vert \cA_{\ell} \vert$. Depending on the selection of $\cI_{\ell}$, the pair of swapped bits diversifies.

\begin{algorithm}\label{Alg:RP}
\caption{Rate-profile}
\KwData{$L$, $(K_{\ell})_{\ell=0}^{L}$, $(N_{\ell})_{\ell=1}^{L}$.}
\KwResult{$\{\mathcal{I}_{\ell}\}_{\ell=0}^{L}$,$\{\mathcal{A}_{\ell}\}_{\ell=1}^{L}$.}
\smallskip

$n_p \gets \sum_{\ell=1}^{L} N_{\ell}$ // required bits for {\it Type-\Rom{1}} pre-transform\;
$\mathcal{R} \defeq \{i_1, \ldots, i_{K_0 + n_p}\}$ // ordered index set in terms of reliability\; 
$w \gets \min_{i \in \mathcal{R}} {\sf wt}({\bf g}_{N, i})$\;
$\cB_{w}(\cR) \gets \{i \in \mathcal{R}: {\sf wt}({\bf g}_{N, i}) = w\}$ // ordered set\;

\medskip
\emph{/* Union of connection set */}\;
$\mathcal{A} \gets \phi$\;
\While{$n_p > 0$}
{   \eIf{$\vert \cB_{w}(\cR) \vert < n_p$}
    {   $\mathcal{A}\gets \mathcal{A}\cup \cB_{w}(\cR)$\;
    } 
    {$\mathcal{A} \gets $ the first $n_p$ elements of $\cB_{w}(\cR)$\;
    }
    $n_p \gets n_p -\vert \cB_{w}(\cR) \vert$\;
    $w \gets 2w$\;
    $\cB_{w}(\cR) \gets \{i \in \mathcal{R}: {\sf wt}({\bf g}_{N, i}) = w\}$ // ordered set\;
}
\medskip
\emph{/* Rate-profile */}\;
$\mathcal{I}_0 \gets \mathcal{R} \backslash \mathcal{A}$\;
$\mathcal{A} \gets {\sf sort}(\mathcal{A}; \text{natural ascending order}) $ \;
\For {$\ell=1$ \KwTo $L$}
{
$\mathcal{A}_{\ell} \gets$ the first $N_{\ell}$ elements of $\mathcal{A}$\;
$\mathcal{A} \gets \mathcal{A} \backslash \mathcal{A}_{\ell}$\;
$\mathcal{I}_{\ell} \gets$ top-$N_{\ell}$ indices with largest ${\sf wt}\left({\bf g}_{i, N_{\ell}}^{\top}\right)$\;
}

Return $\mathcal{I}_{\ell}$ and $\mathcal{A}_{\ell}$\;

\end{algorithm}

\subsection{Type-\Rom{2} Pre-Transform}\label{sec:PT}

\begin{algorithm}
\caption{Design of {\it Type-\Rom{2}} pre-transform}\label{Alg:type-2}
\KwResult{The merged pair $\left(\mathcal{M}_{\ell}\right)$}

\SetKwFunction{Faddpair}{addPair}
\SetKwProg{myproc}{Function}{:}{\KwRet}

$w_{\rm min} \gets \min_{i\in\mathcal{I}_0} {\sf wt}({\bf g}_{N, i})$ \;
$\cB_{w_{\rm min}}(\cI_0) \gets \{ i \in \cI_0 : {\sf wt}({\bf g}_{N, i}) = w_{\rm min} \}$ \;
$\mathcal{I}_{w_{\rm min}} \gets \cB_{w_{\rm min}}(\cI_0)$ \;

$\ell \gets 0$\;
${\sf state} \gets 0$\;
\While{${\sf state} \leq 1$}{
    \For(\tcp*[h]{natural ascend. order}){$i \in \mathcal{I}_{w_{\rm min}}$}{
        
        $\mathcal{P} \gets \left\{ k \in [N] \backslash [i]: k\notin \mathcal{A}_{\ell} \cup \mathcal{I}_{0},~ k \notin \mathcal{M}_{\ell, 2}\right\}$\;  

        \For(\tcp*[h]{natural ascend. order}){$j \in \mathcal{P}$}{
            \If{${\sf state} = 0$ \& ${\sf wt}({\bf g}_{N,j}) \ge w_{\rm min}$} 
            {
            \Faddpair$(i,j)$\;
            {\bf break} \tcp*[h]{break inner for loop} \;
            }          

            
            \If{${\sf state} = 1$ \& ${\sf wt}\left({\bf g}_{N, i} + {\bf g}_{N, j}\right) > w_{\rm min}$} 
            {
            \Faddpair$(i,j)$\;
            {\bf break}\; 
            }
            \If{${\sf state} = 2$ \& ${\sf wt}\left({\bf g}_{N, i} + {\bf g}_{N, j}\right) = w_{\rm min}$} 
            {
            \Faddpair$(i,j)$\;
            {\bf break}\;
            }
        }
    }
    ${\sf state} \gets {\sf state} + 1$\;
    $\mathcal{I}_{w_{\rm min}} \gets \{i\in\mathcal{I}_{w_{\rm min}}: i\notin \mathcal{M}_{\ell, 1}\}$ \;  
}
{\bf Return} $(\mathcal{M}_{\ell})$\;

\myproc{\Faddpair$(i,j)$} {
$\mathcal{M}_{\ell} \gets (i, j)$\; 
$\mathcal{I}_0 \gets \mathcal{I}_0 \backslash \{ i \}$\;
$\ell \gets \ell + 1$\;
}
\end{algorithm}

{\it Type-\Rom{2}} pre-transform aims to reduce $P_{\sf SCL}(\cE_2; S)$ without any loss in $P_{\sf SCL}(\cE_1; S)$. To that end, we group some index $i\in\cI$ with subsequent $j\in\cF$, and apply pre-transform. 
As {\it Type-\Rom{2}} pre-transform matrix, we consider ${\bf G}_2^{\top}$, equivalent to row-merging operation.

To select the pair $(i,j)$, we use Theorem~\ref{thm:1}.
The objective is to merge some information bits with succeeding frozen bits to reduce the number of minimum weight codewords. For description, we introduce the following notations:
\begin{align}
    w_{\rm min}(\cI) &\defeq \min_{i \in \cI} {\sf wt}({\bf g}_{N, i}), \\
    \cB_{w}(\cI) &\defeq \{ i \in \cI : {\sf wt}({\bf g}_{N, i}) = w \}.
\end{align}
Let $\mathcal{M}_{\ell}=(\mathcal{M}_{\ell, 1}, \mathcal{M}_{\ell, 2})$ be the $\ell$th merged-pair, where $\mathcal{M}_{\ell, 1}$ is information index and $\mathcal{M}_{\ell, 2}$ is merged frozen index. 
Given information index $i\in \cB_{w_{\rm min}}(\cI_0)$ and all other previously determined merged-pair $\mathcal{M}_{\ell}$, the candidate of merged index $\mathcal{P}_i$ is given as
\begin{align}
    \mathcal{P}_i \defeq \{ j \in [N] \backslash [i]: j \notin \mathcal{A}_{\ell} \cup \mathcal{I}_0, j \notin \mathcal{M}_{\ell, 2} \},
\end{align}
which collects subsequent frozen indices not belonging to any pre-transform. 

According to Section III-E, the min-weight codewords are generated by the combination of min-weight row, core rows, and balancing rows. To disturb the formation of min-weight codeword, we select merged index $j\in\cP_i$ if 
\begin{align}
    {\sf wt}({\bf g}_{N,j}) \ge w_{\rm min}. \label{eqn:pair-selection-rule-1}
\end{align}
Theorem~\ref{thm:1} ensures that \eqref{eqn:pair-selection-rule-1} eliminates the min-weight codewords in $\cC_i(\cI)$. 
For each index in $i\in\cB_{w_{\rm min}}(\cI_0)$, we take greedy approach assigning the smallest index $j$ satisfying \eqref{eqn:pair-selection-rule-1} to $\cM_{\ell, 2}$.

Next, we repeat the same process using index $i\in (\cB_{w_{\rm min}}(\cI_0) \backslash (\bigcup_{\ell} \cM_{\ell, 1}))$ and takes index $j\in \cP_i$ if
\begin{align}
    {\sf wt}({\bf g}_{N,i} + {\bf g}_{N,j}) > w_{\rm min}. \label{eqn:pair-selection-rule-2}
\end{align}
Again, we select the smallest such index in $\cP_i$. 
Finally, we repeat the same process and takes any takes any index $j\in\cP_i$, which is equivalent to taking index such that 
\begin{align}
    {\sf wt}({\bf g}_{N,i} + {\bf g}_{N,j}) = w_{\rm min}. \label{eqn:pair-selection-rule-3}
\end{align}

We summarize row-merging pair selection process in Algorithm~\ref{Alg:type-2}. To further optimize the row-merged pair, the algorithms enumerating the number of minimum weight codewords can be used such as \cite{zunker-enumeration-pretransform} instead of greedy approach. 

\subsection{Discussion on Type-\Rom{2} Pre-Transform}
To further understand how min-weight codewords are eliminated, we consider the merged pair $(i,j)$. 
The $i$th row ${\bf g}_{N,i}$ devotes to the formation of min-weight codewords in two ways: i) as a coset leader of $\cC_i(\cI \cup \cA)$ and ii) as a core or balancing row of $\cC_{\ell}(\cI \cup \cA)$ with ${\sf wt}({\bf g}_{N, \ell})=w_{\rm min}$, $\ell \in \cI \cup \cA$.

Firstly, suppose the row ${\bf g}_{N,i}$ is a coset leader.  
we divide the merged pair $(i,j)$ into three partitions: 
\begin{itemize}
    \item ${\sf wt}({\bf g}_{N,j}) \ge w_{\rm min}$,
    \item ${\sf wt}({\bf g}_{N,j}) < w_{\rm min} \text{ and }  {\sf wt}({\bf g}_{N,i} + {\bf g}_{N,j}) > {\sf wt}({\bf g}_{N,i})$,
    \item ${\sf wt}({\bf g}_{N,j}) < w_{\rm min} \text{ and }  {\sf wt}({\bf g}_{N,i} + {\bf g}_{N,j}) = {\sf wt}({\bf g}_{N,i})$.
\end{itemize}
If ${\sf wt}({\bf g}_{N, j}) \ge w_{\rm min}$, Theorem~\ref{thm:1} explains the deletion of min-weight codewords in coset $\cC_i(\cI \cup \cA)$. However, when ${\sf wt}({\bf g}_{N, j}) < w_{\rm min}$, we can no longer rely on \eqref{eqn:min-weight-formation}. 
Instead, we utilize the following results in \cite[Theorem 5]{SC-decoding-weight-distribution}. It states that given $i, j \in [0, N-1]$ such that $i < j$, for any ${\bf x} \in \left( {\bf g}_{N, i} + \cC_j(\cI \cup \cA) \right)$, the Hamming weight of generated codewords satisfy the following:
\begin{align}
    {\sf wt} ({\bf x}) \ge {\sf wt} ({\bf g}_{N,i} + {\bf g}_{N,j}).
\end{align}
Because we select the closest subsequent merged index $j$, if ${\sf wt}({\bf g}_{N,i} + {\bf g}_{N,j}) > {\sf wt}({\bf g}_{N,i})$, a large portion of min-weight codewords in $\cC_i(\cI \cup \cA)$ is no longer min-weight codewords. 

Secondly, suppose the row ${\bf g}_{N, i}$ is core or balancing row in coset $\cC_{\ell}(\cI\cup\cA)$ with ${\sf wt}({\bf g}_{N,\ell}) = w_{\rm min}$. Depending on $j$, the inserted row acts as either core row or balancing row. We give explanation based on the conjecture presented in \cite{rowshan-minweight-formation}: i) if row ${\bf g}_{N, j}$ acts as the core row, it might require additional balancing row having lower reliability, which would be frozen row, and ii) if row ${\bf g}_{N, j}$ acts as the balancing row, it would become additional unnecessary balancing row.

\section{Examples}\label{sec:example}

We present some examples illustrating the effect of {\it Type-\Rom{1}} pre-transform, i.e., bit-swapping with pre-transform, and {\it Type-\Rom{2}} pre-transform.

\begin{figure}[t] 
   \centering
   \includegraphics[width=\columnwidth]{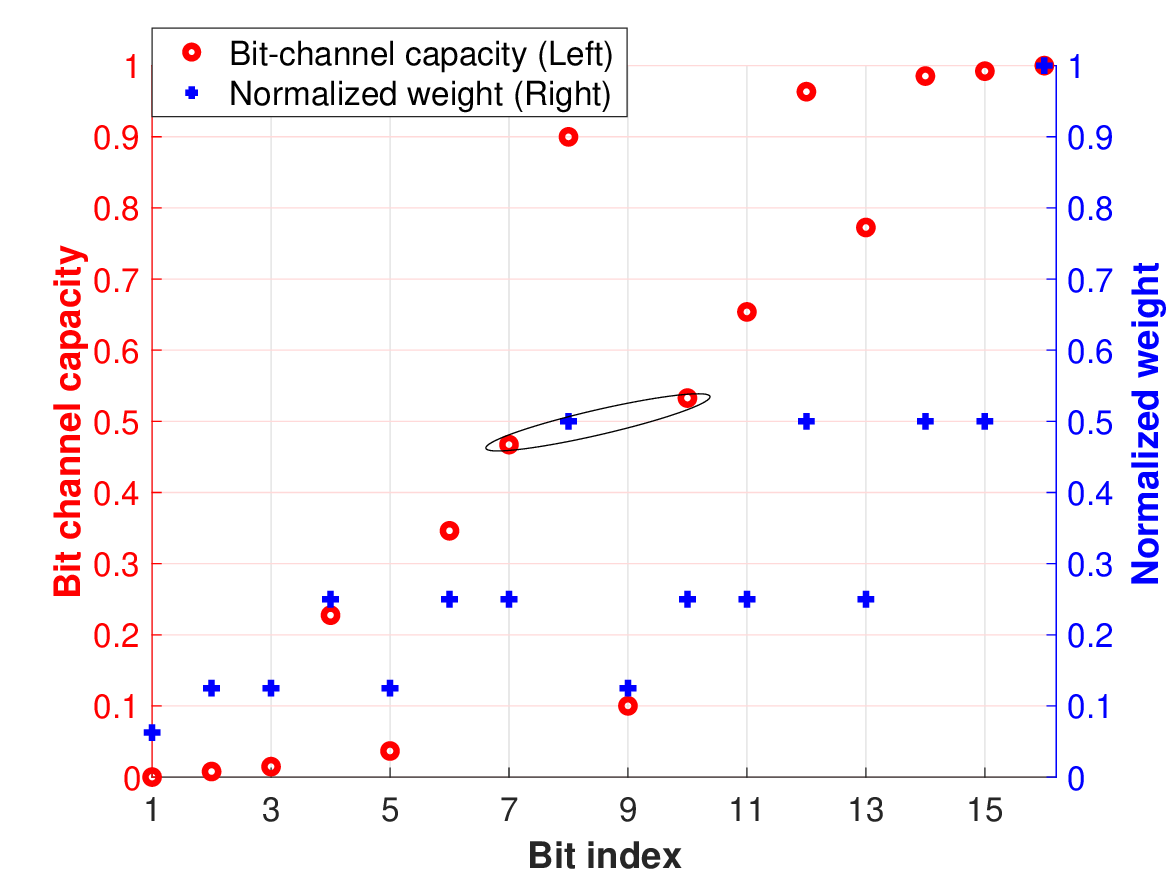} 
   \caption{(Example 1) Bit channel capacity (left) and normalized row weight (right) on binary erasure channel (BEC) with erasure probability of $1/2$.}
   \label{fig:example}
\end{figure}

\vspace{0.2cm}
{\bf Example 1:} 
Consider a short packet transmission scenario, in which a transmitter sends a codeword with a blocklength of $N=16$ over the BEC with an erasure probability of $1/2$, denoted as $I(W_{\sf BEC})=0.5$. The corresponding bit channel capacity and normalized row weight are illustrated in Fig.~\ref{fig:example}. In this example, we present the effect of {\it Type-\Rom{1}} pre-transform. 
Consider a polar code with a code rate of $R = 8/16$. The information index set is given by $\cI_{\sf polar} = \{8, 10, 11, 12, 13, 14, 15, 16\}$, which offers the highest bit-channel capacity. Observe that only possible pre-transform i.e., $u_9 = u_8$, preserves the weight spectrum. 
Now, introduce one additional bits $u_7$, which is the most reliable bit among $\cI_{\sf polar}^c$. Following our rate-profile, we construct connection index set as $\cA_1 = \{7,10\}$, which consists of two most unreliable bits with minimum row weight. Now, if we use pre-transform ${\bf G}_2^{\top}$ with $\cI_1 = \{1\}$ and $\cF_1 = \{2\}$ to maintain a code rate, the number of min-weight codeword $A_{ {\rm d}_{\rm min}}$ reduces from $28$ to $12$ as shown in Table.~\ref{table:example1-WD}.

\begin{table}
\centering
\caption{Comparison of the weight distributions }\label{table:example1-WD}
\begin{tabular}{c|ccccccc}
\toprule
{Weight} & 0 & 4 & 6 & 8 & 10 & 12 & 16 \\ 
\midrule
Polar code & 1 & 28 & - & 198 & - & 28 & 1 \\
RM-type code & 1 & 20 & 32 & 150 & 32 & 20 & 1 \\
SPP code & 1 & 12 & 64 & 102 & 64 & 12 & 1 \\
\bottomrule
\end{tabular}
\end{table}


\vspace{0.2cm}

{\bf Example 2:} 
Recall that SCL decoding error event is partitioned into 
\begin{itemize}
\item $\mathcal{E}_1$: the correct codeword is not in the final list,
\item $\mathcal{E}_2$: the correct codeword is in the final list, while there exists other codewords closer to the received vector.
\end{itemize}
Through the example, we explain that our {\it Type-\Rom{1}} pre-transform  decreases $P(\mathcal{E}_2)$ while nearly maintaining $P(\mathcal{E}_1)$. 
Consider the same scenario with Example 1. 
Because $I\left( W_{16}^{(7)}\right) < I \left( W_{16}^{(10)} \right)$, the SC decoding performance becomes worse due to bit-swapping. If we use SCL decoder, however, the pre-transformed polar code can achieve better performance compared to pure polar code due to improved weight spectrum as shown in Table~\ref{table:example1-WD}. For intuitive explanation, let us assume that SCL decoding with list size of 2 and perfect decoding of ${\hat u}_8$.
Then, considering the decoding of $u_{11}$, decoding path of pure polar code is $({\hat u}_8, {\hat u}_{10}) = \{ ({\hat u}_8, 0), ({\hat u}_8, 1) \}$ and decoding path of pre-transformed polar code is $({\hat u}_7, {\hat u}_8) = \{ (0, {\hat u}_8), (1, {\hat u}_8) \}$. Due to the assumption of perfect estimation of ${\hat u}_8$, the above decoding path contains the every possible combinations of previous bits. From the perspective of the decoding of subsequent bits, in both codes, the behavior of the decoder is almost identical. Therefore, $P(\mathcal{E}_2)$ of pre-transformed polar code is improved while maintaining $P(\mathcal{E}_1)$ under SCL decoding.

\vspace{0.2cm}

{\bf Example 3:}
We use path metric range, introduced in \cite{Rowshan-path-metric-range, Rowshan-bit-swapping}, to observe that non-consecutive information bits produced by {\it Type-\Rom{1}} pre-transform reduces $P(\cE_1)$. 
If we denote the path metric of the $s$th decoding path for the $i$th decoding step as ${\sf PM}_i[s]$, the path metric range for the $i$th decoding step is defined by their maximum difference, 
\begin{equation}
    {\sf PMR}_i = \max_s {\sf PM}_i[s] - \min_s {\sf PM}_i[s].    
\end{equation}
Large path metric range indicates the large difference between the most convincing decoding path and uncertain path in the list, signifying the high potential for the correct decoding path to be discarded.

\begin{figure}[t] 
   \centering
   \includegraphics[width=1.05\columnwidth]{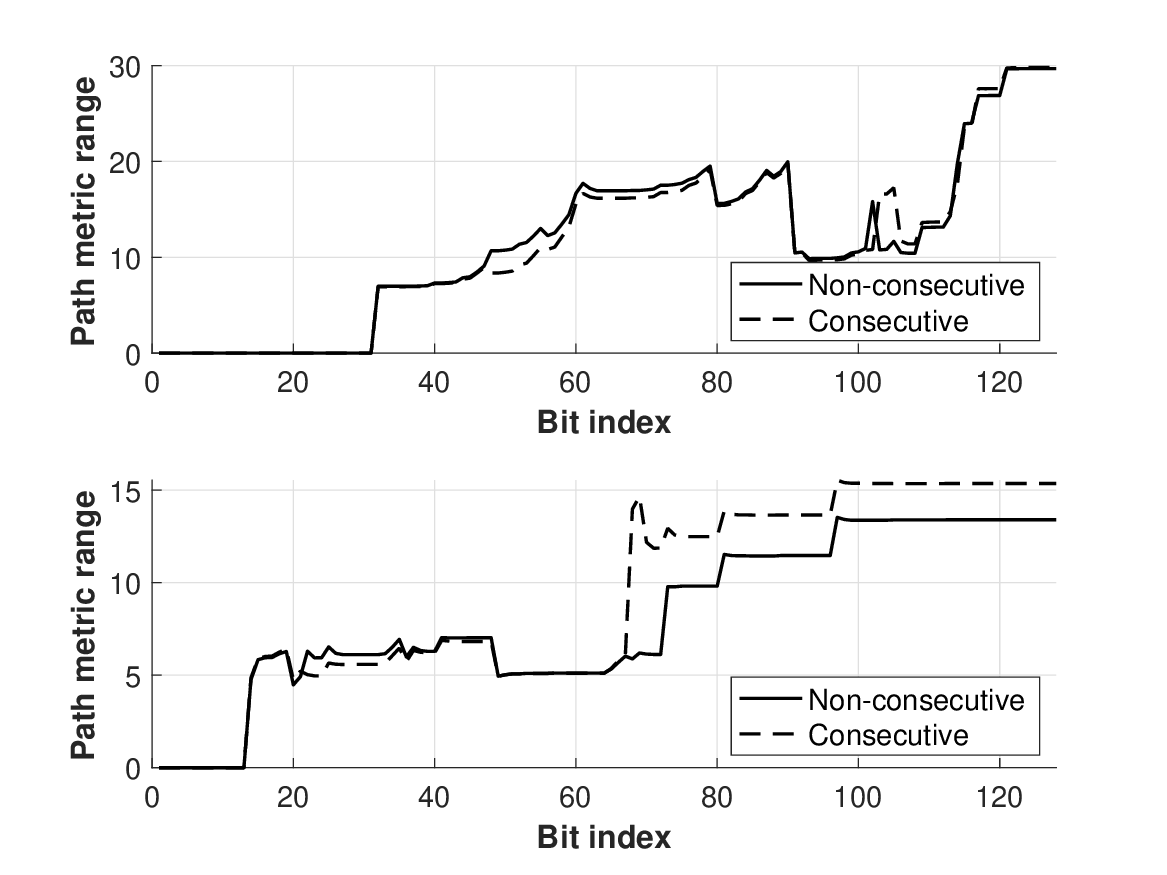} 
   \caption{(Example 3) Path metric range evaluated with list size of $S=2$ averaged over $10^3$ channel noises generated at ${E_b/N_0} = 3$ dB. Top: $(N,K) = (128,32)$, Bottom: $(N,K)=(128,96)$.}
   \label{fig:PMR}
\end{figure}

We present two different {\it Type-\Rom{1}} pre-transform at two different code rates $(N,K)=(128,32)$ and $(N,K)=(128,96)$. First, consider $(N,K)=(128,32)$. The first pre-transform configuration is obtained by parameters $(N_{\ell})_{\ell=1}^{3} = (2, 2, 16)$ and $(K_{\ell})_{\ell=1}^{3} = (1,1,10)$, leading to $\mathcal{A}_1 = \{32, 48\}$, $\mathcal{A}_2 = \{ 56, 60\}$, and $\mathcal{A}_3 = \{ 62, 63, \ldots, 121\}$. It uses 8 additional bits (i.e., 8 bit-swapping), resulting $(d_{\rm min}, A_{d_{\rm min}}) = (24,416)$. The second pre-transform configuration is obtained by $(N_{\ell})_{\ell=1}^{2} = (4, 16)$ and $(K_{\ell})_{\ell=1}^{2} = (3,10)$, resulting in $\mathcal{A}_1 = \{32, 48, 56, 60\}$ and  $\mathcal{A}_2 = \{ 62, 63, \ldots, 121\}$. 
It uses 7 additional bits (i.e., 7 bit-swapping), and $(d_{\rm min}, A_{d_{\rm min}}) = (24,224)$. For the first configuration, when decoding $u_{56}$, because $u_{48}$ is dynamic frozen bit, the gap of path metric between decoding paths is large as shown in Fig.~\ref{fig:PMR} (top). However, for the second configuration, the consecutive semi-polarized bit makes correct decoding path being discarded due to relatively small path metric range, leading to decoding performance loss. 

Consider $(N,K)=(128,96)$. The first configuration is obtained by $(N_{\ell})_{\ell=1}^{3} = (2,2,16)$ and $(K_{\ell})_{\ell=1}^{3} =(1,1,13)$, leading to $\cA_1 = \{14, 15\}, \cA_2 = \{20,22\}$, 
and the second configuration is obtained by $(N_{\ell})_{\ell=1}^{2} = (2,16)$ and $(K_{\ell})_{\ell=1}^{2} =(1,12)$, leading to $\cA_1 = \{14, 15\}, \cA_2 = \{20, 22, 23, 26, \ldots\}$. Both configuration introduce 5 addition bits. 
For the first configuration, when decoding $u_{24}$, because $u_{22}$ is frozen bit, the gap of path metric between decoding paths is large as shown in Fig.~\ref{fig:PMR} (bottom). However, for the second configuration, small path metric range worse the decoding performance when decoded with small list size. 


\vspace{0.2cm}

\begin{figure}[t] 
   \centering
   \includegraphics[width=1.05\columnwidth]{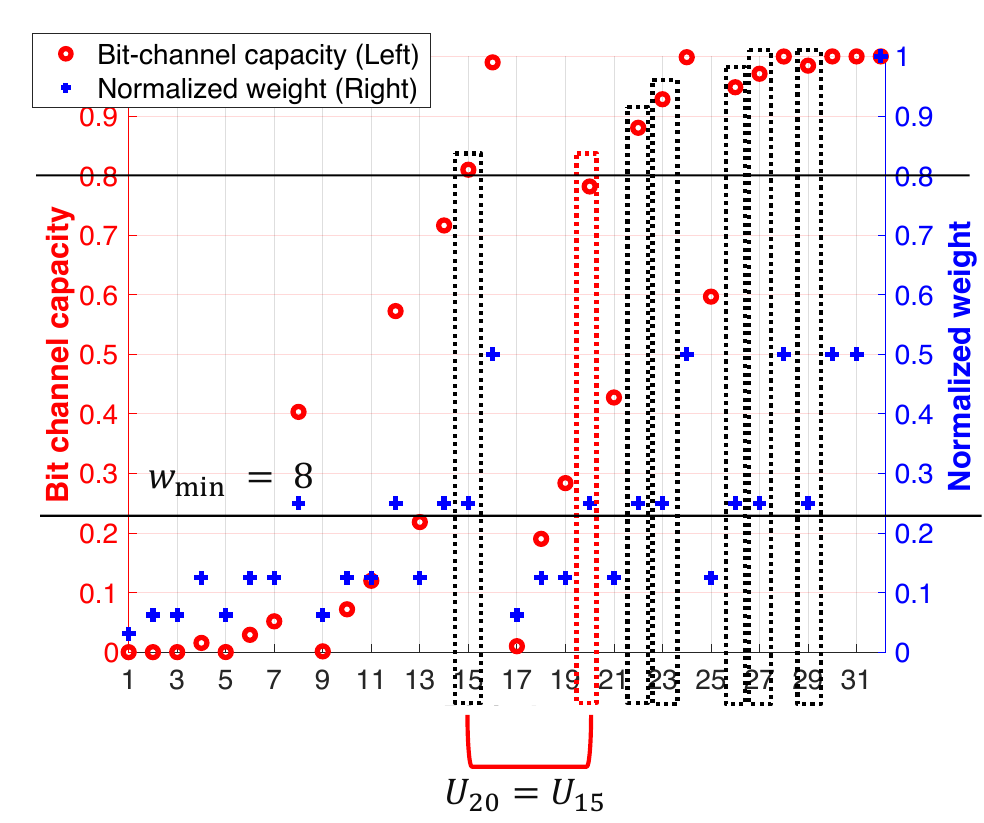} 
   \caption{(Example 4) Bit channel capacity (left) and normalized row weight (right) on binary erasure channel (BEC) with erasure probability of 1/2. }
   \label{fig:example4}
\end{figure}

{\bf Example 4:}
We present example elucidating Theorem~\ref{thm:1}. 
Consider a polar code with blocklength $N=32$, for which the bit channel capacity and normalized row weight is depicted in Fig.~\ref{fig:example4}. 
Define the information set as $\cI = \{ i \in [N]: I\left(W_{N}^{(i)}\right) > 0.8\}$. Observe that a pair of index $(15, 20)$ satisfies the condition of Theorem~\ref{thm:1}. Accordingly, all the min-weight codewords in $\cC_{15}(\cI)$ are deleted, as shown in Table~\ref{table:example4}.

\begin{table}
\centering
\caption{The number of min-weight codewords in $\cC_i(\cI)$}\label{table:example4}
\begin{tabular}{c|ccccccc}
\toprule
{Bit index} & 15 & 22 & 23 & 26 & 27 & 29 & o/w \\ 
\midrule
Before &   32 & 32 & 16 & 16 & 8 & 4 & 0 \\
After &   0 & 32 & 16 & 16 & 8 & 4 & 0 \\
\bottomrule
\end{tabular}
\end{table}

\section{Simulation Results}

We presents the BLER performance of proposed codes under SCL decoding at BI-AWGN channel to evaluate the decoding performance. 


\begin{table*}[t]
\centering
\caption{Construction of SPP Codes} \label{table:simulation-parameters}
\begin{tabular}{c llc ccccc | cc }
\toprule
{} &  {} & {} & {} & \multicolumn{5}{c}{Type-\Rom{1} $(N_{\ell}, K_{\ell})$} &  \multicolumn{2}{c}{Weight} \\
\cmidrule(lr){5-9}\cmidrule(lr){10-11}
{Fig.} & {$R$} & Rate-profile & {CRC}  & {$\ell = 1$} & {$\ell = 2$} & {$\ell = 3$} & {$\ell=4$} & {$\ell=5$} & $d^{\rm min}$ & $A_{d^{\rm min}}$ \\
\midrule
\ref{fig:N128L8} & $1/4$ & 5G & $\phi$ & $(2,1)$ & $(2,1)$ & $(2,1)$ & $(8,3)$ & - & $24$ & $288$ \\
\ref{fig:N128L8} & $2/4$ & 5G & $\phi$ & $(8,2)$ & - & - & - & - & $12$ & $128$ \\
\ref{fig:N128L8} & $3/4$ & 5G & $\phi$ & $(8,3)$ & - & - & - & - & $8$ & $16560$ \\
\midrule
\ref{fig:N256L8} & $1/4$ & 5G & $g_{\sf CRC3}$ & $(4,1)$ & $(4,1)$ & - & - & - & $32$ & $420$ \\
\ref{fig:N256L8} & $2/4$ & 5G & $g_{\sf CRC3}$ & $(4,2)$ & - & - & - & - & $16$ & $1936$ \\
\ref{fig:N256L8} & $3/4$ & 5G & $g_{\sf CRC3}$ & $(8,7)$ & $(8,7)$ & $(8,7)$ & $(8,7)$ & $(8,7)$ & $8$ & $2370$ \\
\bottomrule
\multicolumn{11}{l}{** CRC polynomial: $g_{\sf CRC3}(x) = 1+x+x^3$.} \\
\multicolumn{11}{l}{** Estimate ${d}^{\rm min}$ is computed using SCL decoder with list size $S=10^5$.}
\end{tabular}
\end{table*}

\subsection{Construction of Proposed Codes and Benchmarks}
We consider the following existing codes and benchmarks. 
\begin{itemize}
    \item {\bf CA polar:} The 5G-NR channel-independent reliability sequence \cite{3gpp-nr-coding} and CRC polynomial $g_{\sf CRC11}(x) = x^{11} + x^{10} + x^9 + x^5 + 1$ is used.
    \item {\bf PAC:} 
    We use different rate-profile in each figure. Fig.~\ref{fig:N128L8} is simulated using RM information set, where equality is broken by 5G information set. Fig.~\ref{fig:N256L8} is simulated using 5G information set.
    The convolution polynomial is optimized to reduce the number of min-weight codewords over degree less than $10$. 
    \item {\bf DeepPolar:} A deep polar code is a novel variant of pre-transformed polar codes where the pre-transform consists of multi-layered nested polar encoding \cite{deep-polar-construction-A}. Fig.~\ref{fig:N128L8} is simulated using $(N_{\ell})_{\ell=1}^3 = (2, 16, 128)$ and $(K_{\ell})_{\ell=1}^3 = (1, 10, K-11)$, and Fig.~\ref{fig:N256L8} is simulated using $(N_{\ell})_{\ell=1}^3 = (2, 32, 256)$ and $(K_{\ell})_{\ell=1}^3 = (1, 26, K-27)$.
    \item {\bf SPP:} The 5G-NR information set \cite{3gpp-nr-coding} is used in Fig.~\ref{fig:N128L8} and \ref{fig:N256L8}. In
    The remaining detail of construction is listed in Table~\ref{table:simulation-parameters}.
    \item {\bf Theoretical bounds:} 
     We use two finite blocklength information theoretical bounds: i) random coding union (RCU) bound and ii) meta-converse bound, which are implemented based on \cite{rcu-website} and \cite{github-spectre}. 
    These bounds allow to access the gaps between the achievable decoding performance of our methods and mathematical bounds.
\end{itemize}


\subsection{BLER Comparison}
\begin{figure}[t] 
   \centering
   \includegraphics[width=1.1\columnwidth]{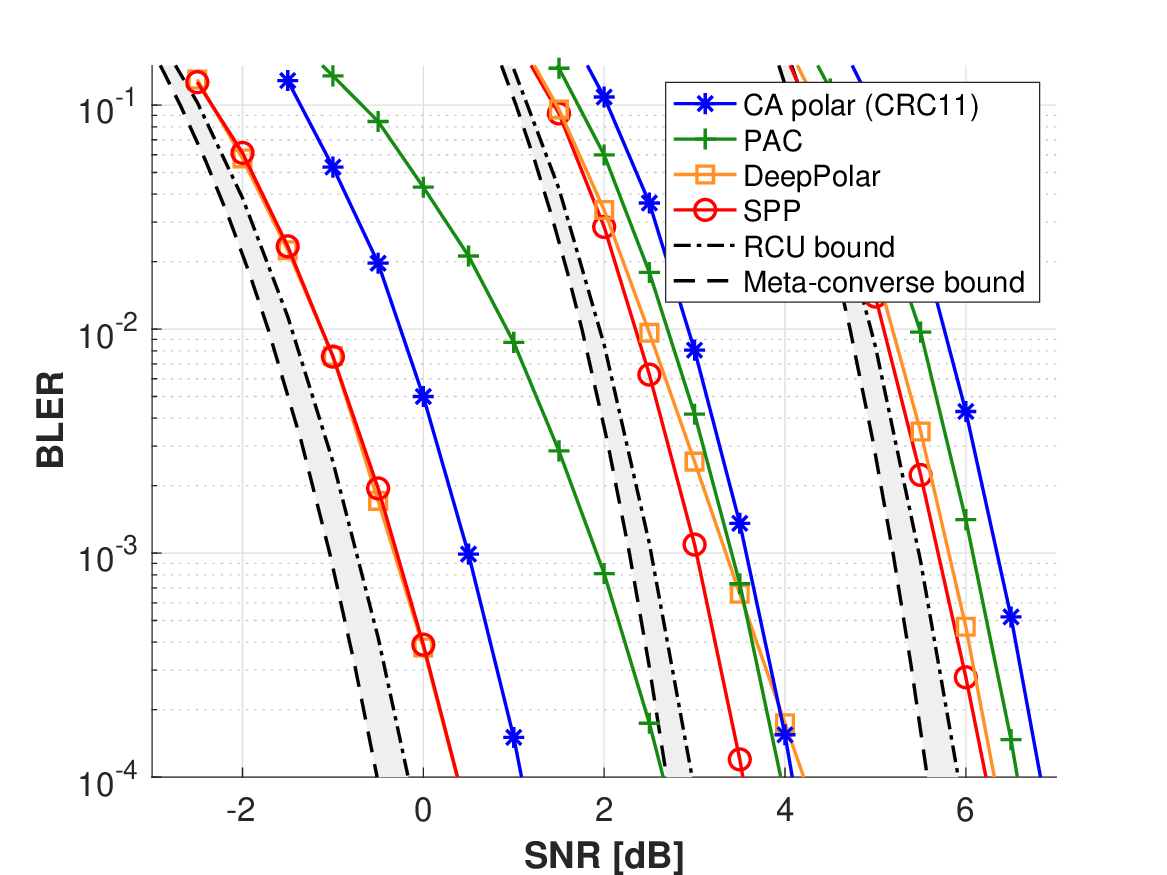} 
   \caption{BLER simulation results using SCL decoder with list size of $S=8$ at blocklength $N=128$.}
   \label{fig:N128L8}
\end{figure}

\begin{figure}[t] 
   \centering
   \includegraphics[width=1.1\columnwidth]{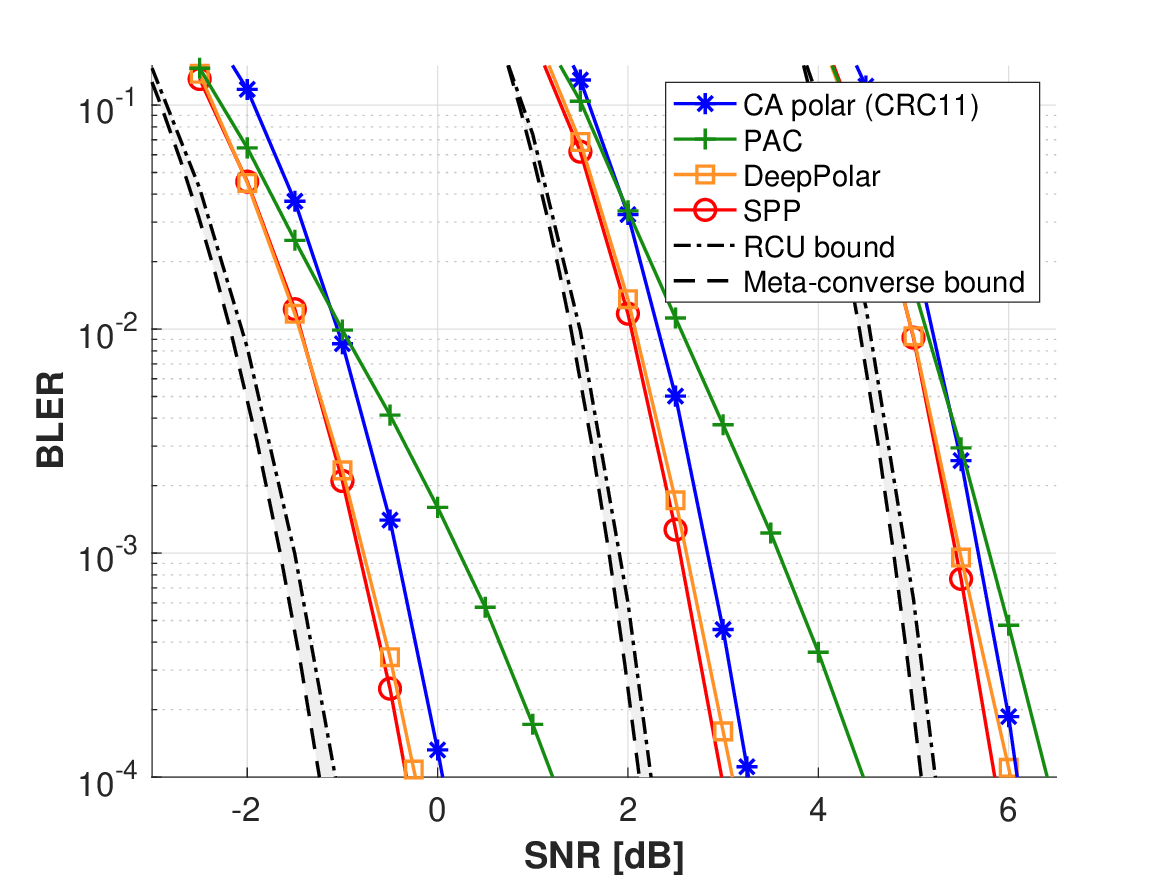} 
   \caption{BLER simulation results using SCL decoder with list size of $S=8$ at blocklength $N=256$.} \label{fig:N256L8}
\end{figure}

{\bf BLER comparison with CA polar codes:}
We consider the blocklength $N\in\{128, 256\}$ and code rate $R\in\{\frac{1}{4}, \frac{1}{2}, \frac{3}{4}\}$. We simulate our proposed and CA-polar codes using an SCL decoder with a list size of $S=8$. Fig.~\ref{fig:N128L8} and Fig.~\ref{fig:N256L8} depict the BLER performance of considered codes at $N=128$ and $N=256$, respectively. The results demonstrate that our method outperforms CA-polar codes at various blocklengths and code rates.
In particular, the sparse pre-transform gain becomes more significant at short blocklength and low rates, even leading to approximately $1$~dB coding gain at blocklength $N=128$ and code rate $R=\frac{1}{4}$. 

{\bf BLER comparison with PAC codes:} Fig.~\ref{fig:N128L8} and Fig.~\ref{fig:N256L8} depict the BLER performance of considered codes at $N=128$ and $N=256$, respectively. The results demonstrate that our method outperforms PAC codes at various blocklengths and code rates.
The decoding performance of PAC codes is inferior due to the RM rate profile (Fig.~\ref{fig:N128L8}), which aims to improve minimum distance, and due to a large number of min-weight codewords (Fig.~\ref{fig:N256L8}) within the 5G information set.
These findings indicate that sparse pre-transform is effective for SCL decoders with small list sizes by limiting the number of consecutive less reliable information bits while simultaneously decreasing the nubmer of min-weight codewords.  


\section{Conclusion}

We have introduced a new type of pre-transformed polar code called SPP, which is an improvement over deep polar codes for low-latency SCL decoding. The main technical innovation involves limiting the number of consecutive semi-polarized information bits while attaining pre-transform gains. To achieve this, we propose applying multiple polar pre-transform matrices in parallel. This parallel pre-transform structure allows for greater flexibility in designing the pre-transform matrix, which enhances both the weight spectrum and decodability under the constraints of small SCL decoding sizes. Based on a quantitative analysis of the SCL decoder's behavior and the formation of min-weight codewords, we designed an algorithm for selecting the connection indices to which pre-transforms are applied. These pre-transforms can be concatenated with global pre-transform techniques, such as CRC precoding, to further improve the distance properties of the resultant codewords. Extensive simulation results under various blocklengths and code rates have demonstrated that our codes consistently outperform all existing state-of-the-art pre-transformed polar codes, achieving superior performance at various rates and short blocklengths while maintaining reasonable decoding complexity.

\bibliographystyle{IEEEtran}
\bibliography{Coding}

\end{document}